\documentclass[11pt,a4paper]{article}

\usepackage{authblk}

\usepackage{amssymb}
\usepackage{amsmath}
\usepackage[dvipsnames]{xcolor}
\usepackage{geometry} 
\usepackage{tikz}
\usetikzlibrary{decorations.pathmorphing,arrows.meta, positioning, calc}

\tikzset{
	bluearrow/.style={<->, thick, blue, >=Stealth},
	redproliferation/.style={<->, thick, red, >=Stealth},
	redproliferation2/.style={->, thick, red, >=Stealth},
	reddestruction/.style={->, thick, red, >=Stealth},
	redinteraction/.style={
		<->, thick, cyan, dashed, >=Stealth,
		decorate,
		decoration={snake, amplitude=2.5pt, segment length=30pt}
	},
	diffusionarrow/.style={->, thick, gray, dashed, >=Stealth},
	chemotaxisarrow/.style={->, thick, green!70!black
		, dashed, >=Stealth},
	nodecircle/.style={circle, draw, minimum size=1.5cm, font=\large, font=\bfseries},
	smallcircle/.style={circle, fill=cyan!60, minimum size=0.25cm, inner sep=0pt}
}

\newcommand{\be}{\begin{equation}}
	\newcommand{\ee}{\end{equation}}

\newcommand{\dis}{\displaystyle}
\newcommand{\pa}{\partial}
\newcommand{\bv}{\mathbf{v}}

\newcommand{\bx}{\mathbf{x}}
\newcommand{\bk}{\mathbf{k}}
\newcommand{\bvp}{\mathbf{v'}}

\newcommand{\nn}{\nonumber}
\newcommand{\fvec}{\underline{\mathbf{f}}}

 \usepackage{amsthm}
 \numberwithin{equation}{section}

 \newtheorem{remark}{Remark}[section]
 \theoremstyle{plain}

 \usepackage{url}

\date{}

\title{Derivation from kinetic theory and 2-D pattern analysis of chemotaxis models for Multiple Sclerosis}

\author[1]{M. Bisi}
\author[1]{M. Groppi}
\author[2]{G. Martal\`o}
\author[3]{R. Travaglini\thanks{Corresponding author. Email: \texttt{travaglini@altamatematica.it}}}

\affil[1]{Dipartimento di Scienze Matematiche, Fisiche e Informatiche,
	Università di Parma, Parco Area delle Scienze 53/A, 43124 Parma, Italy}

\affil[2]{Dipartimento di Matematica ``Felice Casorati'',
	Università di Pavia, Via Ferrata 5, 27100 Pavia, Italy}

\affil[3]{Istituto Nazionale di Alta Matematica ``Francesco Severi'',
	Piazzale Aldo Moro 5, 00185 Roma, Italy}

\date{}

\begin{document}
	
	\maketitle
	
	\begin{abstract}
		In this paper, a class of reaction-diffusion equations for Multiple Sclerosis is presented. 
		These models are derived by means of a diffusive limit starting from a proper kinetic description, 
		taking account of the underlying microscopic interactions among cells. 
		At the macroscopic level, we discuss the necessary conditions for Turing instability phenomena 
		and the formation of two-dimensional patterns, whose shape and stability are investigated 
		by means of a weakly nonlinear analysis. 
		Some numerical simulations, confirming and extending theoretical results, 
		are proposed for a specific scenario.
	\end{abstract}
	
	

	
	
	
	\emph {Declaration of interest:} The authors declare that they have no known competing financial interests or personal relationships that could have appeared to influence the work reported in this paper.

	\section{Introduction}
	
	Multiple Sclerosis (MS) is one of the most severe and debilitating disorders affecting the central nervous system. It is characterized by inflammation of the myelin sheath in the brain, leading to the appearance of focal areas of myelin consumption in the white matter, addressed as lesions or plaques. Myelin is a fatty substance produced in the brain by specialized cells called oligodendrocytes, it surrounds nerve fibers, and acts as an insulator, allowing a quick and efficient transmission of electrical impulses along the nerve cells. Damage caused by MS to both oligodendrocytes and myelin result in progressive physical and neurological disability.
	
	It is mostly accepted that MS originates from an autoimmune response, for which the immune system turns dysfunctional and starts attacking healthy cells, tissues, or organs; specifically, in MS, immune cells such as T-cells, B-cells, macrophages, and microglia (specialized macrophages of the central nervous system) can be activated when matching their cognate antigen expressed by myelin and oligodendrocytes;  { for a comprehensive overview of immune cells involved in MS, we address the reader to the recent review \cite{attfield2022immunology}.} At an early stage of the disease, the patterns of demyelination tend to be similar within each individual but vary significantly between different patients, suggesting the presence of diverse immune mechanisms in plaque formation. Analogously, the clinical progression of MS, the characteristics of lesions, and the resulting irreversible neurological symptoms vary among patients. Researchers have identified four main types of demyelination, classified according to distinct targets of injury and mechanisms of demyelination \cite{lucchinetti2000heterogeneity}. { Although in these findings there was neither observed overlap in pattern nor a transition between different lesion types throughout the clinical progression of individual patients, further studies \cite{de2001post} have individuated a possible correlation between the two most common types of lesions, referred as type II  {  (displayed by around 50\% of patients)} and type III  {  (displayed by around 30\% of patients)}.} More precisely, type III lesions, presenting wide areas of oligodendrocytes injury and activation of microglia with few or no T-cells and absence of remyelination process (restoration of myelin by oligodendrocytes resulting in the formation of ``shadow'' plaques)\cite{attfield2022immunology,prineas2012oligodendrocytes}, have been thought to represent a very early stage (so called-``pre-demyelination'') of the type II lesions, which are indeed characterized by the attack to the myelin sheath by T-cells and abundance of remyelinating shadow plaques.
	For a comprehensive understanding of MS based on medical studies, readers are directed to papers \cite{lassmann2005multiple,lassmann2007immunopathology,lassmann2012progressive,mahad2015pathological,lassmann2022contribution} and references therein.
	
	During each phase of the disease, active demyelination and neurodegeneration are consistently related to inflammation \cite{lassmann2011molecular}, which is widely acknowledged as the primary catalyst for clinical disease and tissue damage \cite{wiendl2009multiple}. A crucial role in inflammation is played by proinflammatory cytokines (chemokines). These
	molecules enter the central nervous system and recruit self-reactive immune cells, that migrate through chemotactic motion \cite{luster1998chemokines}. Once activated, immune cells produce cytokines themselves, attracting other cells to the inflammation site; moreover, cytokines may stimulate clonal expansion of immune cells, enlarging the immune cascade.
	
	{ A crucial role in autoimmune activation is played by families of so-called immunosuppressive cells like regulatory T lymphocytes (Tregs) and natural killer cells. Self-reactive immune cells and cells presenting the activating self-antigen can be found in peripheral tissues even in non-pathological conditions \cite{danke2004autoreactive}, and immunosuppressive cells are, then, able to inhibit or kill them.
		Even if the primary causes of MS are still unknown, a lack of efficiency of immunosuppressive cells is believed to be one of the originating factors, and this weakness of immunosuppressive cells is often reported in MS cases \cite{hoglund2013one,mimpen2020natural,zozulya2008role}.}
	
	In literature, many works have been devoted to the mathematical modeling of MS. Systems of ordinary differential equations have been used to describe relapsing-remitting dynamics \cite{elettreby2020simple,frascoli2022dynamics} or brain damage \cite{kotelnikova2017dynamics}. In \cite{khonsari2007origins}, the classic chemotaxis model by Keller and Segel was adapted to describe the motion of activated microglia via chemotactic signaling of cytokines. This model was refined in \cite{lombardo2017demyelination} and \cite{bisi2024chemotaxis}. In these works, authors pick various specific choices for the chemotactic sensitivity function or production and saturation terms of activated microglia. These models are primarily designed for macroscopic densities of cells and substances, with parameters derived from experimental observations or heuristic considerations. However, the underlying microscopic interactions among cells and molecules are crucial to understanding the observable outcomes. 
	
	The cellular dynamics of immune response exemplifies a complex system composed of numerous heterogeneous living entities, interacting stochastically within themselves and the hosting environment. The kinetic theory of active particles \cite{bellomo2021life} turns out to be a suitable tool for investigating these phenomena,
	in which living organisms interact through sensitivity and visibility mechanisms, related to non-locality and multiple interplays.
	
	The latest results concerning the application of the kinetic theory of active particles to autoimmune diseases are given in \cite{costa2021optimal,della2022mathematical,ramos2019kinetic}, where authors apply such theory to populations of self-antigen presenting cells, self-reactive T cells, and immunosuppressive cells.
	Each population is defined by its microscopic functional state and, through appropriate integration, a macroscopic depiction over time of the behavior of biologically significant quantities can be worked out. Moreover, a description in terms of spatial variables would also replicate immune cell migration, which is related to inflammation and regulated by chemotactic motion induced by cytokines.
	In this perspective, in \cite{Oliveira2022} the authors propose a kinetic description that allows deducing a set of partial differential equations of reaction-diffusion type for autoimmunity. Inspired by works like \cite{othmer1988models,othmer2000diffusion,othmer2002diffusion}, this derivation is achieved through a suitable time scaling, followed by a proper diffusive limit. This approach has been applied in various areas, from classical Boltzmann theory of gases \cite{bisi2006reactive,bisi2022reaction,lachowicz2002microscopic} to the dynamics of cells and tissues (see \cite{burini2019multiscale} and references therein). The same procedure of \cite{Oliveira2022} has been applied for the particular case of MS in \cite{oliveira2024reaction}.
	{ In that work, the authors manage to reproduce 
		patterns mimicking brain lesions characteristic of the usual clinical course of the disease, which usually consists of an initial relapsing-remitting stage, characterized by active myelin lesions and noticeable remyelination, and a secondary progressive phase, during which remyelination becomes less frequent, and other processes contribute to demyelination and neurodegeneration.
		This is obtained by 
		incorporating processes such as myelin sheath consumption by activated immune cells and restoration by oligodendrocytes. The analysis, however, is carried out without focusing on any particular type of lesions, and the formation of lesions is investigated by means of a standard Turing instability analysis.
		
		 { 
		In the present work, we derive a macroscopic system from the kinetic level focusing on the peculiar scenario of the formation of type III lesions, which involve oligodendrocyte lysis induced by activated macrophages. Investigating the underlying dynamics of type III lesion formation may not only help to understand the development of the second most common type of injury in MS, but also offer insights that could aid in preventing the onset of { type II lesions, that represent the most common form.} The modeling objective is to describe the early phases of inflammation and demyelination, focusing specifically on the dynamics between self-antigen presenting cells, immunosuppressive cells, activated microglia, pro-inflammatory cytokines, and oligodendrocytes.
		It is our belief that a mathematical setting for microscopic dynamics would be of interest, as it could lead to a coherent macroscopic scenario for observable phenomena keeping a close connection with the microscopic level.		
		} 
		} Moreover, since some mechanisms are still unknown, we consider a generic shape for functions describing diffusion, chemotactic sensitivity, production, and saturation of microglia. Additionally, to extend results given in \cite{lombardo2017demyelination} and \cite{bisi2024chemotaxis}, where a weakly nonlinear analysis is performed to investigate the emergence of patterns in one dimension, we perform a stability analysis in a two-dimensional domain, following the approach proposed in \cite{Ouyang2000}, showing the formation of different types of patterns. Analytical results are then specified for particular cases, already considered in \cite{lombardo2017demyelination} and \cite{bisi2024chemotaxis}.
	
	The paper is structured as follows: in Section \ref{Sec2} the kinetic setting for the distribution functions of the involved populations is outlined, and the operators accounting for conservative and non-conservative processes are detailed. Under the hypothesis of multiple scale processes, a diffusive limit is performed in Section \ref{Sec3}, to derive a system of reaction-diffusion equations for population densities. The Turing instability analysis of the macroscopic model is presented in Section \ref{Sec4}, providing necessary conditions for the emergence of spatial patterns in a two-dimensional domain; moreover, their shape and stability are discussed through a weakly nonlinear analysis. Numerical simulations are reported in Section \ref{sec5}, in order to confirm the pattern formation predicted by the weakly nonlinear analysis and to investigate the scenarios far from the bifurcation value. Some concluding remarks are given in Section \ref{sec6}.   {The most technical computations are postponed in \ref{Appendix}.}

	\section{Kinetic description}
	\label{Sec2}
	
	The starting point is the description, at the mesoscopic level, of each population involved in the model, along with the different types of evolution dynamics and interactions occurring among them. Inspired by \cite{ramos2019kinetic}, 
	we consider self-antigen presenting cells ($A$) and immunosuppressive cells ($S$); then, instead of self-reactive  T-cells, we take into account self-reactive microglia ($M$). Moreover, as done in previous works \cite{Oliveira2022,oliveira2024reaction}, we include the cytokines population ($C$). Lastly, we add the oligodendrocytes population,  dividing them into three subgroups: healthy ($D_1$), attacked  ($D_2$), and destroyed ones ($D$).
	
	The behavior of each population is described by a proper kinetic equation for its own distribution function.
	Distributions will depend on time $t\in \mathbb R_0^+$ and space $\bx \in \Gamma_{\bx}$, with $\Gamma_{\bx}$ a bounded domain in $\mathbb R^2$. In addition, we consider the activity variable $u\in[0,1]$ for cell populations, i.e. all populations except cytokines. The activity variable represents the amount of activation of each cell with respect to its specific role (see \cite{ramos2019kinetic} for further details).  {  Accordingly to the immunology of MS, the disease originates when self-reactive immune cells, immunosuppressive cells and antigen presenting cells manage to migrate into the central nervous system \cite{attfield2022immunology,moise2021mathematical}, triggering the autoimmune cascade. In the present work, though, we want to focus on the dynamics taking place in the white matter, which involve microglia, cytokines and oligodendrocytes (this distinction is also adopted in the model proposed in \cite{moise2021mathematical}). For this reason we shall neglect the motion of  immunosuppressive cells  and antigen-presenting cells.} For microglia, instead, the distribution function also depends on the velocity variable, 
	in order to include spatial diffusion and chemotaxis interplay; thus, we consider microglia velocity $\bv\in \Gamma_{M}=\mathcal V\mathbb B$,
	with $\mathcal V$ the maximal speed
	and $\mathbb {B}$ the unit ball in $\mathbb{R}^2$. 
	 {  On the other hand, oligodendrocytes do not migrate, since they extend processes along axons to form myelin sheaths \cite{tiane2019opc}.} Thus, we consider distribution functions $f_I(t, \bx, u)$, $I=A,\,S,\,D_1,\,D_2,\,D$, $f_M(t,\bx,\bv,u)$. 	
	Macroscopic densities depending on time and space are obtained as moments of the distribution functions, by integrating them with respect to activity and/or velocity
	 { 
	\be \nn
	\begin{aligned}
		\dis I(t,\bx) &= \int_0^1 f_I(t, \bx, u)\,du,\quad I=A,S,D_1,D_2,D,
		\\[3mm]
		\dis  \rho_M(t, \bx, u)&=\int_{\Gamma_M} f_M(t,\bx,\bv,u)\,d\bv,\quad \dis M(t,\bx)=\int_0^1 \rho_M(t, \bx, u)\,du.
	\end{aligned}
	\ee
}

	Each distribution function is governed by an integro-differential equation. More precisely, the evolution for $A$, $S$, $D_1$, $D_2$ and $D$ is described by the following equation 
	\be
	\dis\frac{\pa f_I}{\pa t}=   \mathcal G_I(\fvec)+ \mathcal N_I(\fvec)+ \mathcal I_I(\fvec),\quad I=A,\,S,\,D_1,\,D_2,\,D,
	\label{eq:kin1}
	\ee
	whereas for $M$, whose distribution depends also on the velocity variable, also a drift term appears
	\be
	\dis \frac{\pa f_M}{\pa t}+\bv\cdot\nabla_{\bx}\, f_M= \mathcal G_M(\fvec)+ \mathcal {L}_M(f_M)+ \mathcal N_M(\fvec)+ \mathcal I_M(\fvec),
	\label{eq:kin2}
	\ee
	where we indicate by $\fvec$ the vector of all distribution functions. 
	 {  For cytokines, instead,  {   due to their molecular nature and not being self-propelling entities, we neglect the velocity in $f_C(t,\bx)$ and,}	being their migration in space several orders  of magnitude larger that the speed of the microglia \cite{moise2021mathematical}, we consider directly a diffusion term for the macroscopic density $C(t,\bx)$, writing
	\be
	\dis \frac{\pa C}{\pa t}+D_C\,\Delta_{\bx}\, C= \mathcal N_C(\fvec)+ \mathcal I_C(\fvec),
	\label{eq:kinC}
	\ee
	being $D_C$ the diffusion coefficient.
	}
	The right-hand sides of \eqref{eq:kin1}, \eqref{eq:kin2}, and \eqref{eq:kinC} contain the terms accounting for interactions with other agents or with the external environment. In detail, the terms $ \mathcal G_I$  and $\mathcal {L}_I$ are proper integral operators related to the outcome of conservative processes, i.e. those interplays whose result is only a change in the activity or in the velocity of agents. The terms of type $ \mathcal N_I$, instead, describe the role of binary interactions among agents, that may be proliferative or destructive for the population $I$. Terms of type $ \mathcal I_I$, finally, collect the proliferation or destruction effects which depend on other processes. Interactions and the corresponding operators will be listed in the following subsections.
	
	\subsection{Conservative interactions}

	We adopt the same hypotheses of \cite{ramos2019kinetic,Oliveira2022} and we suppose that binary interactions among self-antigen presenting cells, microglia, and immunosuppressive cells induce a change (increase or decrease) in the activity of each participating cell  { (for a more detailed biological justification of the performed choices, we address the reader to \cite{ramos2019kinetic}). 
		More specifically, interactions are listed as follows.
		\begin{itemize}
			\item[-]  The interactions between self-antigen presenting cells and macrophages enhance the formers' activity by increasing their ability to activate macrophages. This, in turn, enhances macrophages' functional state, allowing them to more effectively recognize self-antigens as foreign agents, 
			\be A+M\rightarrow A^++M^+ \label{consAR},\ee
			indicating from now on through the index $+$ ($-$) the fact that, as a result of the interaction, the activity is increased (decreased);
			\item[-]  Interactions between self-antigen presenting cells and immunosuppressive cells reduce the ability of the former to activate macrophages, while the latter's ability to inhibit the autoimmune response decreases after the interactions,
			\be A+S\rightarrow A^-+S^- \label{consAS};\ee
			\item[-]  Macrophages engage in conservative interactions with immunosuppressive cells, in which their ability to activate and produce cytokine is weakened due to the inhibitory effect of immunosuppressive cells, and also in this case the latter's activity decreases after the interactions,
			\be M+S\rightarrow M^-+S^- \label{consRS}.\ee
		\end{itemize}
		 {  
		The corresponding conservative operators can be defined as done in \cite{ramos2019kinetic}, where authors outline the functions $\eta_{IJ}(v,w)$, that account for the interaction rates between a cell of population $I$ having activity $v$ and a cell of population $J$ having activity $w$, and functions 
		$\mathcal C_{IJ}(v,w;u)$ that represent the transition probability for a cell of population $I$ having activity $v$ to pass to activity $u$ after the interaction with a cell of population $J$ having activity $w$. Thus, conservative operators read as follows.
	}
		\begin{align}
			\nn\mathcal G_A&(\fvec)=\,
			\\ \nn
			&
			\int_{\Gamma_M} \int_0^1 \int_0^1 \eta_{A M}(u^*, u') \, \mathcal{C}_{A M}(u^*, u'; u) \, f_A(t,\bx,u^*) \, f_M (t,\bx,\bv,u') \, du^* \, du'\,{d\bv} 
			\\ \nn
			& - f_A(t,\bx,u)\int_{\Gamma_M} \int_0^1 \eta_{A M}(u, u') \, f_M (t,\bx,\bv,u') \, du'\,{d\bv}\\ \nn
			& +\int_0^1 \int_0^1 \eta_{A S}(u^*, u') \, \mathcal{C}_{A S}(u^*, u'; u) \, f_A(t,\bx,u^*) \, f_S (t,\bx,u') \, du^* \, du' 
			\\ 
			& - f_A(t,\bx,u) \int_0^1 \eta_{A S}(u, u') \, f_S (t,\bx,u') \, du',\label{eq:ga}
		\end{align}
		\begin{align}
			\nn\mathcal G_M&(\fvec)=\,\\ \nn
			&
			\int_{\Gamma_M} \int_0^1 \int_0^1 \eta_{M A}(u^*, u') \, \mathcal{C}_{MA}(u^*, u'; u) \, f_M (t,\bx,\bv,u^*)\, f_A (t,\bx,\bv,u') \, du^* \, du'\,{d\bv} 
			\\ \nn
			& - f_M(t,\bx,\bv,u) \int_0^1 \eta_{MA}(u, u') \, f_A(t,\bx,u') \, du'\\ \nn
			& + \int_{\Gamma_M} \int_0^1 \int_0^1 \eta_{M S}(u^*, u') \, \mathcal{C}_{M S}(u^*, u'; u) \, f_M(t,\bx,\bv,u^*) \, f_S (t,\bx,u') \, du^* \, du' 
			\\ 
			& - f_M(t,\bx,\bv,u) \int_0^1 \eta_{MS}(u, u') \, f_S (t,\bx,u') \, du',\label{eq:gm}
		\end{align}
		\begin{align}
			\nn\mathcal G_S&(\fvec)=\, 
			\\ \nn
			&\int_0^1 \int_0^1 \eta_{S A}(u^*, u') \, \mathcal{C}_{SA}(u^*, u'; u) \, f_S(t,\bx,u^*) \, f_A (t,\bx,u') \, du^* \, du' 
			\\ 
			& - f_S(t,\bx,u) \int_0^1 \eta_{SA}(u, u') \, f_A (t,\bx,u') \, du'
			\\ \nn
			& +\int_{\Gamma_M} \int_0^1 \int_0^1 \eta_{S M}(u^*, u') \, \mathcal{C}_{S M}(u^*, u'; u) \, f_S(t,\bx,u^*) \, f_M (t,\bx,\bv,u') \, du^* \, du'\,{d\bv} 
			\\ \nn
			& - f_S(t,\bx,u)\int_{\Gamma_M} \int_0^1 \eta_{S M}(u, u') \, f_M (t,\bx,\bv,u') \, du'\,{d\bv} ,\label{eq:gs}
		\end{align}
			\be\label{etaCi1}\eta_{IJ}(v,w):=c_{IJ}(v-1)^2,\quad \mathcal C_{IJ}(v,w;u)=\frac{2(u-v)}{(v-1)^2}\mathbf 1_{u>v},\mbox{ for} (I,J)\in\{(A,M),(M,A)\},\ee
			\be\label{etaCi2}\eta_{IJ}(v,w):=c_{IJ}(v)^2,\quad \mathcal C_{IJ}(v,w;u)=\frac{2(u-v)}{(v)^2}\mathbf 1_{u<v},\mbox{ for} (I,J)\in\{(A,S),(M,S),(S,M),(S,A)\},\ee
			where coefficients $c_{IJ}$ are positive constants.{  The shape of functions adopted above are inspired by the choices made in \cite{ramos2019kinetic}, which represent particular cases of a more general kinetic theory framework originally developed to model tumor–immune system interactions \cite{bellomo1994dynamics}. In this setting, the interaction rates $\eta_{IJ}$ depend solely on the activity level of the cell in population $I$, and increase either quadratically with the difference between the activity and its maximal value (as in \eqref{etaCi1}) or with the activity itself (as in \eqref{etaCi2}). The transition probabilities $\mathcal{C}_{IJ}$ also depend on the pre- and post-interaction activity levels of the cell in population $I$, and are proportional to their difference, capturing the assumption that activity can only increase (as in \eqref{etaCi1}) or decrease (as in \eqref{etaCi2}). Finally, normalization condition $\int_0^1
				 \mathcal{C}_{IJ} \, du = 1$ holds.
			\begin{remark}			
			 We observe that transition probabilities and collision kernels ensure the conservative nature of operators $ \mathcal{G}_I$, since they are such that $\int_0^1 \mathcal{G}_I du=0$, $I=A,M,S$, which means that no change in the total number of interacting agents occur. { We show that this property holds for the first two terms on the right-hand side of \eqref{eq:ga}:
		 $$
		 \begin{aligned}
		 	&\int_0^1\left(
		 	\iint_{\tilde\Gamma} \int_0^1 \eta(u^*, u') \, \mathcal{C}(u^*, u'; u) \, f_A(u^*) \, f_M (u') \, du^* \, du'\,{d\bv}  - f_A(u)\iint_{\tilde\Gamma} \eta(u, u') \, f_M (u') \, du'\,{d\bv}\right)du=
 \\[3mm]&
 \int_0^1\left[
\iint_{\tilde\Gamma} \left( \int_0^u c \,2 (u-u^*) \, f_A(u^*) \, f_M (u') \, du^* \right)\, du'\,{d\bv}  - f_A(u)\iint_{\tilde\Gamma} c(u-1)^2 \, f_M (u') \, du'\,{d\bv}\right]du=
\\[3mm]&
c\,M\int_0^1\left(
  \int_0^u\,2 (u-u^*) \, f_A(u^*) \, du^* - f_A(u)(u-1)^2 \right)du=
\\[3mm]&
c\,M\left[\left(\left.u^2\int_0^u f_A(u^*) \, du^*-u\int_0^u 2 u^* f_A(u^*)\, du^*\right)\right|_{u=0}^{u=1}-
\int_0^1 f_A(u)(1-2u) du
 \right]=0.
 \end{aligned}
$$
For brevity, we have omitted the dependence on $t$, $\bx$, $\bv$ and the sub-index $AM$ in $\eta_{AM}$ and in $c_{AM}$, and we have defined $\tilde\Gamma=\Gamma_M\times[0,1]$.
	For the remaining terms the procedure is analogous.
}
\end{remark}
}		
		Among the conservative processes, we consider also the movement of microglia and cytokines in the environment. At the mesoscopic level, this is described by changes in velocity regulated by an integral turning operator, relying on velocity-jump processes. We suppose that the change in velocity may be random for both microglia and cytokines, but we add an external bias for microglia representing chemotactic attraction due to cytokines, able to influence the movement of cells. 
		{ We suppose that the movement of cells is of a run-and-tumble type, i.e. it alternates straight-line movements (runs) and random (or biased) reorientations (tumble). This dynamics is usually described by a velocity jump process \cite{stroock1974some,alt1980biased,loy2020kinetic,hillen2006m5,conte2023mathematical}. The bias represented by the chemotactic attraction will be described by means of a perturbation of a symmetric probability of the velocity, as performed in classical works modeling chemotaxis \cite{othmer1988models,othmer2000diffusion,othmer2002diffusion}.}
		 {  
			Thus, the turning operator for microglia reads as 
			\be 
			\mathcal {L}_M[C](f_M)(\bv) = \mathcal {L}_M^0(f_M)(\bv)+\mathcal {L}_M^1[C](f_M)(\bv).
			\label{eq:lm}
			\ee
			 As proposed in \cite{oliveira2024reaction}, the probability of a 
			cell to pass from velocity $\bvp$ to $\bv$ is expressed through the uniform probability over the space of velocities, with $\omega=\pi\,\mathcal V^2$, while the turning rate is mediated by a function of the macroscopic density of microglia $\varphi_0(M)$. This is described by the term 
			\be 
			\label{L0m}
			\mathcal {L}_M^0(f_M)(\bv) =  \frac{\lambda}{\varphi_0(M)} \left(-f_M(\bv) + \frac{1}{\omega}\int_{\Gamma_{M}} f_M(\bvp) d\bvp\right).
			\ee			
			 Beside this, the reorientation of the cell towards the cytokines gradient is given by the term
			 \be 
			 \label{L1m} \mathcal {L}_M^1[C](f_M)(\bv) = \gamma\int_{\Gamma_{M}} T^1_M(\bv,\bv',C)f_M(\bvp) d\bvp,
			 \ee
			 in which the turning kernel $ T_M^1 $ is 
			 \be\label{T1m}
			 T^1_M(\bv,\bv',C)=\varphi_1(M)\hat{\bv}\cdot\hat{\bvp}(\hat{\bvp}\cdot\nabla_{\bf x} C) \frac{v}{\mathcal V},
			 \ee   
			 with  $\bv=v \hat{\bv}$, $|\hat{\bv}|=1$. It works  as follows: when $ \hat{\mathbf{v}}' \cdot \nabla_{\mathbf{x}} C(t, \mathbf{x}) > 0 $, $ T_M^1 $  reaches its maximum value when $ \hat{\mathbf{v}} = \hat{\mathbf{v}}' $; conversely, when $ \hat{\mathbf{v}}' \cdot \nabla_{\mathbf{x}} C(t, \mathbf{x}) < 0 $, it attains its maximum when $ \hat{\mathbf{v}} = -\hat{\mathbf{v}}' $. This forces the cells to move in a direction that is aligned with $ \nabla_{\mathbf{x}} C(t, \mathbf{x}) $. At the same time, the increase in the cell speed is highly expected, being the probability proportional to $v$.}
		The term $\gamma\,\varphi_1(M)$, with $\gamma$ positive constant, represents the chemotactic sensitivity. We point out that several choices may be considered for functions $\varphi_0$ and  $\varphi_1$, depending on which phenomenon is taken into account, e.g. the ``volume-filling" effect \cite{wang2007classical}.  {  This choice for the kernel is inspired by one of those firstly proposed in \cite{othmer2002diffusion}, in which the external signal bias determining the change in the cell direction is based on the alignment between the bias gradient and the incoming direction. Along with the chosen scaling, it will provide, at macroscopic level, the classical chemotaxis term related to the Patlak–Keller–Segel–Alt model \cite{othmer2002diffusion}.}
		
		\begin{remark}
			Operators $\mathcal {L}_M^0(f_M)$, $\mathcal {L}_M^1[C](f_M)$
			 satisfy the spectral properties required for the derivation of {a} reaction-diffusion macroscopic model and ensure the conservativity of operators $\mathcal {L}_M^0(f_M)$, $\mathcal {L}_M^1[C](f_M)$, being their integral over the variable $\bv$ null. For more general results and proofs, we address readers to classical references \cite{othmer2000diffusion,othmer2002diffusion}.
		\end{remark}
		
		\subsection{Nonconservative interactions}
		
		As anticipated above, the interactions among cells can lead to proliferative or destructive phenomena  \cite{ramos2019kinetic,Oliveira2022}. In particular, we consider the following proliferative dynamics (also here, we  refer the reader to \cite{ramos2019kinetic} for a broader view of the biological mechanisms modeled here):
		\begin{itemize}
			\item[-] interactions between self-antigen presenting cells and microglia may lead to proliferation for both populations, while interactions between 
			self-antigen presenting cells and immunosuppressive cells may lead to the proliferation of the latter 
			\be A+M\rightarrow A+A+M \label{AAM},\ee
			\be A+M\rightarrow A+M+M \label{AMM},\ee
			\be A+S\rightarrow A+S+S \label{ASS},\ee
			in any case, the newborn cell inherits the same activity as its mother cell;
			\item[-] interactions between self-antigen presenting cells and microglia stimulate microglia to produce cytokines
			\be A+M\rightarrow A+M+C \label{AMC}.\ee
		\end{itemize}
		On the other hand, we include the following destructive processes:
		\begin{itemize}
			\item[-] immunosuppressive cells $S$ cell induce apoptosis (programmed cell death) of both $A$ and $M$ cells
			\be A+S\rightarrow S \label{AS},\ee
			\be M+S\rightarrow S \label{MS}.\ee
			\item[-] microglia attack oligodendrocytes: we distinguish two different phases of the phagocytosis process \cite{zajicek1992interactions,dhib2007pathogenesis}, thus we have an initial  adherence to healthy oligodendrocyte, which turns into an attacked one
			\be M+D_1\rightarrow M+D_2 \label{MD12},\ee
			and then we have a second killing and final phagocytosis phase, resulting in the destruction of the oligodendrocyte
			\be M+D_2\rightarrow M+D \label{MD2D}.\ee			
		\end{itemize}
		
		{Thus we can write the nonconservative operators for $A$, $M$, $S$, $C$, $D_1$, $D_2$, $D$ population accounting for processes \eqref{AAM}-\eqref{MD2D}.  
			We obtain
			 {
			\begin{align} \nn
				\mathcal N_A&(\fvec)=\,
				  f_A(t,\bx,u) \int_{\Gamma_{M}} \! \int_{0}^1 p_{AM}(u,w) f_M(t,\bx,\bv,w) d w d\bv
				\\[2mm]\nn	& \qquad\qquad   -  f_A(t,\bx,u)  \int_{0}^1d_{AS}(u,w)  f_S (t,\bx,w) d w,
				\\[2mm]\nn
				\mathcal N_S&(\fvec)=\,  f_S(t,\bx,u) \int_{0}^1\, p_{SA}(u,w)  f_A(t,\bx,w) d w,
				\\[2mm]\nn
				\mathcal N_M&(\fvec)= \,\pi(M)\,  f_M(t,\bx,\bv,u) \int_{0}^1 \,p_{MA}(u,w) f_A(t,\bx,w) d w 
				\,\\[2mm]\nn
				& \qquad\qquad - \,\pi(M)\, f_M(t,\bx,\bv,u)  \int_{0}^1 d_{MS}(u,w) f_S (t,\bx,w) d w,
				\\[2mm]\nn
				\mathcal N_C&(\fvec)=  \! \int_{\Gamma_{M}} \! \int_0^1  \! \int_0^1 \!\!\,q_{AM}(u,w)
				f_A(t,\bx,u) f_M(t,\bx,\bv,w) du dw d\bv,
				\\[2mm]\nn
				\mathcal N_{D}&(\fvec) =\,
				 f_{D_2}(t,\bx,u)   \int_{\Gamma_{M}}   \int_0^1  \,b_{2M}(u,w)  f_M(t,\bx,\bv,w)  dw d\bv,
				\\[2mm]\nn
				\mathcal N_{D_1}&(\fvec) =\, -  f_{D_1}(t,\bx,u)   \int_{\Gamma_{M}}   \int_0^1\,  b_{1M}(u,w) f_M(t,\bx,\bv,w)  dw d\bv,
				\\[2mm]\nn
				\mathcal N_{D_2}&(\fvec) =\,
				-\,  {\mathcal N}_{D}(\fvec) - {\mathcal N}_{D_1}(\fvec).
			\end{align} 
			To our aims, in the following we will take coefficients $p_{IJ}$, $d_{IJ}$, $q_{IJ}$ and $b_{IJ}$ as positive constants.} As for the proliferative processes, we suppose that newborn cells inherit the same activity of their mother cells. Moreover, we suppose that proliferation and suppression rates for microglia, deriving from interactions with antigen-presenting cells and immunosuppressive cells, respectively, also depend on the macroscopic density of $M$ through the function $\pi(M)>0$.\\

			\subsection{Operators corresponding to other processes}
			We include in the description the natural death of { self-antigen presenting cell and immunosuppressive cell populations and decay of cytokines, occurring at constant rate $d_I$, with $I=A,\,S,\,C$.} Moreover, we take into account the process introduced in \cite{della2022mathematical}, i.e. a constant input of self-antigen presenting cells, depending on
			external factors, which we indicate by $\alpha$. For cytokines, we consider, in addition, the production of the chemical signal by the oligodendrocytes, as proposed in \cite{lombardo2017demyelination,khonsari2007origins} and characterized by the constant rate $q_C$.
			Lastly,  {  despite being the interplay between microglia and oligodendrocytes still under investigation \cite{aydinli2022two}, and} since some studies suggest that both oligodendrocyte injury and the first stage of microglia-induced apoptosis are, in general, reversible \cite{scolding1989reversible,hornik2016activated,zakharov2020problem}, we also consider  the process
			\be \label{rem2} D_2\rightarrow D_1,\ee 
			with constant coefficient $r_1$.
			{The} 
			operators accounting for these processes are
			 {
			\begin{align}\nn                
				\mathcal I_A(\fvec)=\,&\alpha - d_A(u)  f_A(t,\bx,u),\qquad   
				\mathcal I_S(\fvec)=\,  - d_S(u)  f_S(t,\bx,u) \\[2mm]\nn
				\mathcal I_C(\fvec) =\,& { - d_C C(t,\bx)+\,\int_0^1q_{C}(u)\,f_D(t,\bx,u)du}
				\\[2mm] \nn
				\mathcal I_{D_1}(\fvec) = \,&  { \,\int_0^1r_{1}(u)\,f_{D_2}(t,\bx,u) du,} \qquad 
				\mathcal I_{D_2}(\fvec) =\, 
				- \mathcal I_{D_1}(\fvec),
			\end{align} and also  coefficients  $d_{I}$, $q_{C}$ and $r_{1}$ will be taken from now on as positive constants.}

			 { 
			All the populations and parameters in the kinetic model are listed in Table \ref{TabKin}, whereas 	 {  in Figure \ref{schema} are graphically schematized the interactions listed in the kinetic description of the model.}	
			\begin{table}[ht]
				\centering
					\caption{Populations and parameters involved in the kinetic description}\label{TabKin}
				\begin{tabular}{|c|l|}
					\hline
					$A$ & Self-antigen presenting cells \\
					$S$ & Immunosuppressive cells \\
					$M$ & Self-reactive microglia \\
					$C$ & Cytokines \\
					$D_1$ & Healthy oligodendrocytes \\
					$D_2$ & Attacked oligodendrocytes \\
					$D$ & Destroyed oligodendrocytes \\
					 {$c_{IJ}$}&	 { Coefficients of the conservative interaction rates}\\ 
					$p_{IJ}$ & Proliferation rates due to interactions between populations \\
					$d_{IJ}$ & Apoptosis (death) rates due to interactions \\
					$q_{AM},\,q_C$ & Cytokine production rates from interactions and destroyed oligodendrocytes \\
					$b_{KM}, K=1,2$ & Oligodendrocyte damage rates in the two phases of phagocytosis \\
					$d_I$ & Natural death rate of population $I = A,\,S,\,C$ \\
					$\alpha$ & External input rate of self-antigen presenting cells \\
					$r_1$ & Recovery rate from attacked ($D_2$) to healthy ($D_1$) oligodendrocytes \\
			    	$\gamma$ & Chemotactic sensitivity constant\\
			    	$\lambda$ & Rate coefficient in microglial turning frequency \\
					$\mathcal{V}$ & Maximum magnitude of microglial velocity \\
					$\omega$ & Measure of velocity space \\
					\hline
				\end{tabular}
			\end{table}}		
		\begin{figure}
			\centering
		\begin{tikzpicture}[node distance=2cm and 3cm]
		
		\node[nodecircle, draw=cyan!30, fill=cyan!30 ,text=black] (A) {A};
		\node[nodecircle, draw=orange!30, fill=orange!30, right=of A,text=black] (M) {M};
		\node[nodecircle, draw=purple!30, fill=purple!30, below=of A,text=black] (S) {S};
		\node[nodecircle, draw=green!30, fill=green!30, below=of M,text=black] (D) {$\mathbf{\bar{D}}$};
		\node[nodecircle, right=1.3cm of D,white, text=black] (C) {C}; 

		\draw[bluearrow, bend left=25] (A) to node[above, font=\large] {$+$} (M);
		\draw[bluearrow, bend right=25] (A) to node[left, font=\large] {$-$} (S);
		\draw[bluearrow, bend left=25] (M) to node[above, font=\large] {$-$} (S);
		
		\draw[redproliferation, bend right=40] (A) to node[below] {{\large $+$}} (M);
		\draw[redproliferation2, bend left=40] (A) to node[right] {{\large $+$}} (S);
		
		\draw[reddestruction, bend right=30] (S) to node[below left] {{\large $-$}} (A);
		\draw[reddestruction, bend right=30] (S) to node[below right] {{\large $-$}} (M);
		
		\draw[redinteraction] (A) -- node[above] {} (M);
		
		\draw[redinteraction,->] (M) -- node[left] {{\large $+$}} (C);
		
		\draw[redinteraction,->] (D) -- node[above] {{\large $+$}} (C);
		
		\foreach \angle in {0,45,90,135,225,315} {
			\draw[diffusionarrow] 
			($ (M.center) + ( {0.75*cos(\angle)}, {0.75*sin(\angle)} ) $) -- 
			++( {cos(\angle)}, {sin(\angle)} );
		}
		
		\foreach \angle in {0,45,90,135,225,270,315} {
			\draw[diffusionarrow] 
			($ (C.center) + ( {0.5*cos(\angle)}, {0.5*sin(\angle)} ) $) -- 
			++( {cos(\angle)}, {sin(\angle)} );
		}
		
		\draw[chemotaxisarrow, bend left=35] (M) to node[above right] {CHEM.} (C);
		
		\draw[redproliferation, loop above, looseness=10] (A) to node[above, font=\normalsize] {{\large $+$} $\alpha$} (A);
		
		\draw[reddestruction] (M) -- node[left] {{\large $-$}} (D);
		
		\foreach \x/\y in {0.25/0, -0.25/0, 0/0.25, 0/-0.25, 0.18/0.18, -0.18/0.18, 0.18/-0.18, -0.18/-0.18, 0.35/0.15, -0.35/-0.15} {
			\node[smallcircle] at ($ (C.center) + (1.5*\x, 1.8*\y) $) {};
		}
		
	\end{tikzpicture}
	\caption{Quantities and processes involved in the kinetic description. Nodes represent self-antigen presenting cells ($A$), immunosuppressive cells ($S$), self-reactive microglia ($M$), cytokines  ($C$) and total oligodendrocytes ($\bar D$). Conservative processes \eqref{consAR}-\eqref{consAS}-\eqref{consRS} (blue arrows), proliferative processes \eqref{AAM}-\eqref{AMM}-\eqref{ASS}, destructive processes \eqref{AS}-\eqref{MS}, oligodendrocytes destruction by microglia, constant input of self-antigen presenting cells, and destructive processes  (red arrows), interaction \eqref{AMC} and cytokines production (dashed light-blue arrows), chemotaxis motion (dashed green arrow), and diffusion (dashed gray arrows).}\label{schema}
		\end{figure}

			\section{Diffusive limit}
			\label{Sec3}
			
			In this section, our aim is to apply asymptotic methods to obtain a diffusive limit {of the kinetic system \eqref{eq:kin1}-\eqref{eq:kinC}}, 
			as commonly done in kinetic theory for different scenarios in gas dynamics  \cite{lachowicz2002microscopic,bisi2006reactive,bisi2022reaction}, 
			and already applied to active particles \cite{bellomo2009complexity},
			and cells \cite{alt1997dynamics,
				bellomo2004class}. 
			The basic assumption is to suppose that various processes occur at different time scales. For this reason, by a suitable non-dimensionalization, we can put in evidence a small characteristic parameter $\varepsilon$ and set the following temporal hierarchy:
			\begin{enumerate}
				\item velocity-jump processes are the quickest ones, thus the contribution $\mathcal{L}_M$ is of order $\varepsilon^{-1}$;
				\item the reorientation of microglia towards cytokines gradient is supposed to occur at a slower rate (of magnitude $\varepsilon$) with respect to the random movement. This can be expressed as
				\be 
				\mathcal {L}_M[C](f_M)(\bv) = \mathcal {L}_M^0(f_M)(\bv)+\varepsilon\,\mathcal {L}_M^1[C(t,\bx)](f_M)(\bv) ;
				\label{eq:lmeps}
				\ee
				\item  conservative and non-conservative interactions and all the remaining processes constitute the slowest dynamics. In particular,  {  since, as stated before, we are focusing  on the dynamics involving microglia, cytokines and oligodendrocytes,} we find convenient to distinguish two slow scales: processes relevant to populations $A$ and $S$ are of order $\varepsilon^2$, while processes for $M$ and $C$ are of order $\varepsilon$;
				\item  finally, we make the assumption that dynamics \eqref{MD12}  and \eqref{rem2} are slower (order $\varepsilon^2$), than \eqref{MD2D} (order $\varepsilon$); this assumption is based on the fact that recent studies assert that macrophages induce maturation of oligodendrocytes and that mature oligodendrocytes apoptosis lasts more days \cite{aydinli2022two,chapman2024oligodendrocyte}.
			\end{enumerate}
			Setting the time scale of order $\varepsilon$,  {  say $t'=\varepsilon t$ and omitting the apex $'$ for a lighter notation,} from \eqref{eq:kin1}-\eqref{eq:kinC}, we obtain the following rescaled kinetic system
				\begin{align}
					\dis \varepsilon & \,\frac{\pa f_A}{\pa t} 
					= \varepsilon^2\,\mathcal  G_A(\fvec)+ \varepsilon^2\,\mathcal  N_A(\fvec)+ \varepsilon^2\,\mathcal I_A(\fvec) ,
					\label{BolA}
					\\
					\dis \varepsilon  &  \,\frac{\pa f_S}{\pa t}
					= \varepsilon^2\,\mathcal  G_S(\fvec)+ \varepsilon^2\,\mathcal  N_S(\fvec) + \varepsilon^2\,\mathcal I_S(\fvec) ,
					\label{BolS} \\
					\dis \varepsilon & \,\frac{\pa f_M}{\pa t} + \bv\cdot\nabla_{\bx}\, f_M
					= \frac1\varepsilon \mathcal {L}_M[C](f_M) + \varepsilon\,\mathcal  N_M(\fvec)
					\label{BolR}
					\\
					\dis \varepsilon & \, \frac{\pa C}{\pa t}+ \varepsilon D_C\,\Delta_{\bx} C
					=  \varepsilon \, \mathcal { N_C}(\fvec) + \varepsilon\,\mathcal I_C(\fvec),
					\label{BolC}
					\\
					\dis \varepsilon & \, \frac{\pa f_{D_1}}{\pa t}
					= \varepsilon^2 \, {\mathcal N}_{D_1}(\fvec) + \varepsilon^2\,{\mathcal I}_{D_1}(\fvec),
					\label{BolE1}
					\\
					\dis \varepsilon & \, \frac{\pa f_{D_2}}{\pa t}
					= -\varepsilon^2\,{\mathcal N}_{D_1}(\fvec) 
					-\varepsilon^2\, {\mathcal I}_{D_1}(\fvec)-\varepsilon \,  {\mathcal N}_{D}(\fvec) \label{BolE2}
					,
					\\
					\dis \varepsilon & \, \frac{\pa f_{D}}{\pa t}
					= \varepsilon \, {\mathcal N}_{D}(\fvec),			
				\end{align}
				with $
				\mathcal {L}_M[C](f_M)(\bv)$ as in \eqref{eq:lmeps}.
				It can be easily observed that the total number of oligodendrocytes is preserved. 
				
				Following the procedure proposed in previously cited papers for different physical and biological settings, we consider the  { Hilbert} expansion of each distribution function in powers of $\varepsilon$ \cite{bellomo2004class}, i.e.
				$f_I ={ f_I^0} + \varepsilon f_I^1 + \varepsilon^2f_I^2 + O(\varepsilon^3),$ for $I=A,\,S,\,M,\,D_1,\,D_2,\,D$. Without loss of generality, following the framework in \cite{othmer2000diffusion}, we assume that, for $k \geq 1$,
				\be\nn
				\int_0^1 f_I^k(t, \bx, u)\,du=0, \ \  I=A,\,S,\,D_1,\,D_2,\,D,\quad
				\int_{\Gamma_M}f_M^k(t, \bx,\bv,u)\,d\bv=0.
				\ee
				 {  These assumption imply that the total mass for each population is concentrated in the first term of the expansion and that the the remaining terms are relevant only to the changes in the activity and/or in the velocity. Moreover, terms $\epsilon^k f_I^k,k\geq1$ are small perturbations of $f_I^0$ and thus they are allowed to be negative without losing physical consistency.}
				
				We start by considering equations {\eqref{BolA} and \eqref{BolS} for} the populations $A$ and $S$, respectively.
				Inserting expansions for the distribution functions and
				collecting the same order terms in $\varepsilon$, we get 
				\begin{align}
					\dis  \,\frac{\pa f_A^0}{\pa t} &
					=0,\qquad \frac{\pa f_A^1}{\pa t}= \mathcal  G_A[f_A^0,f_M^0,f_S^0]+\mathcal  N_A[f_A^0,f_M^0,f_S^0]+ \mathcal I_A(f_A^0),
					\label{BolAze}
					\\
					\dis  \,\frac{\pa f_S^0}{\pa t} &
					=0,\qquad   \frac{\pa f_S^1}{\pa t}= \mathcal  G_S[f_A^0,f_M^0,f_S^0]+\mathcal  N_S[f_A^0,f_M^0,f_S^0]+ \mathcal I_S(f_S^0).
					\label{BolSze}
				\end{align}
				Then, by integrating the equations above with respect to the activity variable $u$, we may write the following relations (omitting here and in the following $O(\varepsilon)$ terms){
				\begin{equation}
					A(t,\bx)=\Lambda ,\qquad
					S(t,\bx)=  \frac{p_{AM}}{d_{AS}}\,M(t,\bx)-\Sigma\quad\mbox{ with }\quad\Lambda=\frac{d_S}{p_{SA}},\quad\Sigma=\frac{d_A d_S-\alpha\, p_{SA}}{d_S\, d_{AS}} \label{ASMac}
					,
				\end{equation}}
				 { Therefore the density of cells $A$ is constant, while the evolution of population $S$ will follow from that of $M$.}
				
				Now we consider equation \eqref{BolR} for microglia.
				Equating terms of the same order in $\varepsilon$, we 
				obtain
				\begin{eqnarray}
					-\mbox{ order } \varepsilon^0:& & \mathcal {L}_M^0(f_M^{0}) = 0,
					\label{eq:e0}
					\\[2mm]
					-\mbox{ order } \varepsilon^1:& & \bv\cdot\nabla_{\bx}\, f_M^0 = \mathcal {L}_M^0(f_M^{1})+\,\mathcal {L}_M^1[C](f_M^{0}), 
					\label{eq:e1} 
					\\[2mm]
					-\mbox{ order } \varepsilon^2: & & \nn  	\lefteqn{\frac{\pa f_M^0}{\pa t} \!+\! \bv\!\cdot\!\nabla_{\bx} f_M^1 	=  \mathcal {L}_M^0(f_M^{2})}
					\\
					& & \qquad\qquad\qquad\quad
					\!+\!\, \mathcal {L}_M^1[C](f_M^{1})
					\!+\! \mathcal N_M(f_A^0,f_M^0,f_S^0) 
					\label{eq:e2} 
				\end{eqnarray}
				As shown in previous works where the same technique has been adopted \cite{Oliveira2022,oliveira2024reaction,conte2023mathematical}, the spectral properties of the operator $\mathcal {L}_M^0(f_M)(\bv)$ allow us to write
				\be \label{f0rho}
				\begin{aligned}
				f_M^0{ (t,\bx,\bv,u)} &= \rho_M(t,\bx,u) ,\\
				f_M^{1} { (t,\bx,\bv,u)} 
				&= - \frac{\varphi_0(M)}{\lambda} \bv \cdot \nabla_{\bx} \rho_M
				+  \rho_M\int_{\Gamma_{M}} T_M^1(\bv,\bvp,C) d\bvp,
	     		\end{aligned}
				\ee
				with $T_M^1(\bv,\bvp,C)$ defined in \eqref{T1m}. By inserting the terms {$f_M^0$ and $f_M^1$} in equation \eqref{eq:e2}, 
				the term $ f_M^2$ may be recovered by imposing the proper solvability condition (namely that the integral with respect to $\bv$ over the domain $\Gamma_M$ vanishes), which leads to
				\begin{align}
					\nn \frac{\pa \rho_M}{\pa t}& - \nabla_{\bx}\,\cdot  \left[D_M\,\varphi_0(M)\,\nabla_{\bx}\,\,\rho_M -\chi\,\varphi_1(M)\,\rho_M \,\nabla_{\bx}\, \,C\right]
					\\
					&  = \mathcal N_M(f_A^0,\rho_M,f_S^0) 
					\\
					&  = { \pi(M)\, \rho_M \left(p_{MA}\,\frac{d_S}{p_{SA}}-d_{MS}\left(\frac{\alpha\, p_{SA}-d_A d_S}{d_S\, d_{AS}} + \frac{p_{AM}}{d_{AS}} M\right)\right)
					},
					\label{eq:rorr}
				\end{align}
				where we have obtained the diffusion coefficient $D_M$ and the chemotactic parameter $\chi$ as
				\be \label{DRChi}
				D_M = \frac{\mathcal V^2}{4\,\lambda} \quad \mbox{ and} \quad 
				\chi = \frac{\gamma\,\pi\mathcal V^3}{8},
				\ee
				respectively. By integrating also with respect to the activity variable $u$ and relying on relations \eqref{ASMac}, together with \eqref{f0rho}, we end up with
				\begin{equation}
					\label{eqMmac}
					\begin{aligned}
						\frac{\pa M(t,\bx)}{\pa t} &=
						\nabla_{\bx}\,\cdot\left[D_M\,\varphi_0(M(t,\bx))\nabla_{\bx}\,\,M(t,\bx) \right.\\&\qquad\left.- \chi \,\varphi_1(M(t,\bx))\,M(t,\bx)  \,\nabla_{\bx}\,C\right]\\[1mm] &
						\qquad + \pi\left(M(t,\bx)\,\right)M(t,\bx) \left(\psi-\zeta M(t,\bx)\right)
						,
					\end{aligned}
				\end{equation}
				with{
				\be\label{etazeta}
				\psi={{p_{MA}}\Lambda+d_{MS}\Sigma,\quad \zeta={\frac{p_{AM}d_{MS}}{d_{AS}}}.}
				\ee	}
				
				 { For cytokines density, we straightforwardly get
				 the reaction-diffusion equation	
				\begin{equation}
					\frac{\pa C(t,\bx)}{\pa t}=D_C\, \Delta_{\bx}\,C(t,\bx) + b\, M(\bx,t)- d_C C(\bx,t)+ q_C D(\bx,t) ,
					\label{eqCmac}
				\end{equation}
				with {
				$
				b={q_{AM}\,\Lambda}.
				$}
			}
			
				We now deduce evolution equations for the oligodendrocytes. The Hilbert expansion applied to equation \eqref{BolE1} and \eqref{BolE2} provides
				\begin{align}
				\dis  \, \frac{\pa f^0_{D_1}}{\pa t}
				&=0,\quad
				\frac{\pa f^1_{D_1}}{\pa t}
				= \, {\mathcal N}_{D_1}[\rho_M,f^0_{D_1}] + \,{\mathcal I}_{D_1}[f_{D_2}^0],\label{E1cost_a}
				\\\dis  \, \frac{\pa f^0_{D_2}}{\pa t}
				&=-\,{\mathcal N}_{D}[\rho_M,f^0_{D_2}] 
			    ,\\
				\frac{\pa f^1_{D_2}}{\pa t}
				&= \, -\left({\mathcal N}_{D_1}[\rho_M,f^0_{D_1}] + \,{\mathcal I}_{D_1}[f_{D_2}^0]\right)- \mathcal N_D[\rho_M, f_{D_2}^1],\label{E1cost}
				\end{align}
				that lead to 
				\be r_1 \,{D_2}(t,\bx) - b_{1M}{D_1}(t,\bx) M(t,\bx)=0,\label{relE1E2}\ee
				and
				\be
				\dis  \, \frac{\pa {D_2}(t,\bx)}{\pa t}=
				\, -b_{2M}{D_2}(t,\bx)\,M(t,\bx)
				.\label{eq:E2}
				\ee
				{ We suppose that macroscopic oligodendrocyte density, which results in being constant in time, is also constant in space, and we define 
					\be
					D_1(t,\bx)+D_2(t,\bx)+D(t,\bx)=\bar D.
					\ee Thus,} observing that to the leading order ${\pa_t D}=-{\pa_t D_2}$ from \eqref{E1cost_a}, and using \eqref{relE1E2}, we may write down the equation for destroyed oligodendrocytes
				\begin{align}
					\frac{\pa D(t,\bx)}{\pa t}&=(\bar D - D(t,\bx))\frac{b_{2M}\,M(t,\bx)}{\mu+M(t,\bx)}M(t,\bx),\,\mbox{ with}\mu=\frac{r_1}{b_{1M}}.
					\label{eqDmac}
				\end{align}
			
			 { 
			At this point, we collect equations  for each population of the model, obtaining{
			\be
			\begin{aligned}\label{SistMacFull}
				A(t,\bx)=&\,\Lambda, \\
				S(t,\bx)=&\, \frac{p_{AM}}{d_{AS}}\,M(t,\bx)-\Sigma,\\
				\frac{\pa M(t,\bx)}{\pa t} =&
				\nabla_{\bx}\,\cdot\left[D_M\,\varphi_0(M(t,\bx))\nabla_{\bx}\,\,M(t,\bx) - \chi \,\varphi_1(M(t,\bx))\,M(t,\bx)  \,\nabla_{\bx}\,C\right]\\ &
				\qquad + \pi\left(M(t,\bx)\,\right)M(t,\bx) \left(\psi-\zeta M(t,\bx)\right),\\
				\frac{\pa C(t,\bx)}{\pa t}=&\,D_C\, \Delta_{\bx}\,C(t,\bx) + b\, M(\bx,t)- d_C C(\bx,t)+ q_C D(\bx,t),\\
				\frac{\pa D(t,\bx)}{\pa t}=&\,(\bar D - D(t,\bx))\frac{b_{2M}\,M(t,\bx)}{\mu+M(t,\bx)}M(t,\bx).
			\end{aligned}
			\ee}
			} {We  non-adimensionalize the system, adopting the change of variables 
				\be\nn \widetilde t=\psi\, t,\qquad \widetilde{\bx}=\sqrt{\frac{\psi}{D_M}}\bx.\ee 
				Then, we introduce the non-dimensionalized quantities as follows{
				\be \nn
				\widetilde A=\frac{A}{\Lambda},\quad
				\widetilde S=\left(\Lambda\frac{p_{MA}}{d_{MS}}+\Sigma\right)^{-1}\,{S},\quad
				\widetilde M=\frac{\zeta}{\psi}\,{M},\quad
				\widetilde C=\frac{\zeta\,d_C}{\psi\,\bar b}\,{C},\quad
				\widetilde D=\frac{D}{\bar D}.
				\label{eq:ndimdens}
				\ee}
				Defining the new coefficients of the model {as}
				\be \nn
				\xi=\chi\,\frac{\bar b}{D_M d_C\,},\quad
				\tau=\frac{\psi}{d_C},\quad
				\theta=\frac{\psi\,D_C}{d_C\,D_M},\quad
				\delta=\frac{\zeta\,q_C\,\bar D}{\psi\,\bar b},\ee
				\be 
				\beta=\frac{b}{\bar b},\quad
				r=\frac{b_{2M}}{\zeta},\quad\nu=\frac{\mu\,\zeta}{\psi},
				\label{CoefAdim}
				\ee
				and functions 
				\be\nn
				\tilde\varphi_k(x)=\varphi_k\left(x\,\frac{\psi}{\zeta}\right),\quad k=0,1,\quad	\tilde\pi(x)=\pi\left(x\,\frac{\psi}{\zeta}\right)
				\ee
				we get the {following dimensionless equations}{
				\begin{align}
					\label{MacComp1}A=\,&1,
					\\[2mm]
			    	\label{MacComp2}S=\,&\Theta+M,
					\\[2mm]
					\frac{\pa M}{\pa t}=\,&\nabla_{\bx}\cdot\left(\Phi_0(M)\,\nabla_{\bx}\, M-{\xi} \,\Phi_1(M)\,\nabla_{\bx}\,C\right)+\Pi(M) , \\[2mm]
				     \frac{\pa C}{\pa t}=\,&\frac{1}{\tau}\left(\theta\Delta_{\bx}\,C-C+\beta\,M+\delta\,D\right)
					,\\[2mm]
					 \label{MacComp5}\frac{\pa D}{\pa t}=\,&r\,\left(1-D\right) \Psi(M),
				\end{align}}
				where we have renamed the non-dimensional densities by removing the tilde and we have defined
				\begin{equation}
					\begin{aligned}
						&{\Theta=-\Sigma\left(\Lambda\frac{p_{MA}}{d_{MS}}+\Sigma\right)^{-1}},\\
						&\Phi_0(M)=	\tilde\varphi_0(M),\quad\Phi_1(M)=\tilde\varphi_1(M)M,\\ &\Pi(M)=\tilde\pi(M)M(1-M),\quad\Psi(M)=\frac{M}{\nu+M}\,M .
					\end{aligned}
			\end{equation}}
			 {  From the biological point of view, the macroscopic functions here derived provide the modeling of the diverse processes. The function $\Phi_0(M)$ describes
			the diffusivity of the cells due to unbiased (random) movement, while $\Phi_1(M)/M$ is the chemotactic sensitivity that determines the advective flux related to
			the gradient of the signal \cite{hillen2009user}. The term $\Pi(M)$ accounts for the microglia growth, in particular the part $M(1-M)$ models the logistic growth, while the function $\tilde\pi(M)$ includes other processes and we suppose that it does not vanish. Lastly, the function $\Psi(M)$ describes the action of microglia in damaging oligodendrocytes \cite{khonsari2007origins}.			
				We point out that system \eqref{eq:rdsR}-\eqref{eq:rdsE} provides the derivation from the kinetic level of a generalized form of PDEs systems proposed in the literature to describe the formation of type III lesions in MS \cite{lombardo2017demyelination,bilotta2018eckhaus,bisi2024chemotaxis,barresi2016wavefront}, allowing us to relate the coefficients of the macroscopic model to the biological microscopic dynamics. In particular, those models are recovered by taking $\tilde\varphi_0(y)=1$, $\tilde\varphi_1(y)=\left(1+y\right)^{-1}$. These choices correspond to assuming a constant diffusivity and a chemotactic sensitivity function that accounts for the prevention of overcrowding, also known as the ``volume-filling" effect. Moreover, by taking $\tilde\pi(y)=\left(\mu\left(y-h\right)\right)^{i-1},$ (with $\mu>0$ and $h<1$), choosing $i=1$, one gets logistic growth, while setting $i=2$ the Allee effect, which is a growth function used in population
				dynamics to take into account undercrowding effects \cite{stephens1999allee}, is included.						
			}
			{
			The first two equations yield algebraic relations and hence decouple from the dynamics. Therefore, for subsequent analysis we retain only the three evolution equations for $	M,  C,\text{ and } D,$
			as they fully capture the nontrivial spatio‑temporal dynamics, focusing on the system
				\begin{align}
				\label{eq:rdsR} \frac{\pa M}{\pa t}=\,&\nabla_{\bx}\cdot\left(\Phi_0(M)\,\nabla_{\bx}\, M-{\xi} \,\Phi_1(M)\,\nabla_{\bx}\,C\right)+\Pi(M) , \\[2mm]
				\label{eq:rdsC}\frac{\pa C}{\pa t}=\,&\frac{1}{\tau}\left(\theta\Delta_{\bx}\,C-C+\beta\,M+\delta\,D\right)
				,\\[2mm]
				\label{eq:rdsE} \frac{\pa D}{\pa t}=\,&r\,\left(1-D\right) \Psi(M).
			\end{align}
		}For the reader's convenience, all the populations and parameters in the macroscopic model are listed in Table \ref{TabMac}.
		\begin{table}[ht]
		\centering
		\caption{Populations and parameters involved in the macroscopic description \eqref{eq:rdsR} - \eqref{eq:rdsE}}\label{TabMac}
		\begin{tabular}{|c|l|}
			\hline
			$M$ & Self-reactive microglia \\
			$C$ & Cytokines \\
			$D$ & Destroyed oligodendrocytes  \\
			$\xi$ & Chemotactic sensitivity coefficient \\
			$\tau$ & Time scale of cytokine dynamics \\
			$\theta$ & Cytokine diffusion coefficient \\
			$\beta$ & Cytokine production rate by microglia \\
			$\delta$ & Cytokine production rate by damaged oligodendrocytes \\
			$r$ & Damaging intensity of microglia on oligodendrocytes \\
			\hline
		\end{tabular}
		\end{table}				
		 {Concerning rigorous results on existence of solutions for the macroscopic equations, we address the reader to the discussion reported in \cite{gargano2024cytokine}, in which a particular extension of system \eqref{eq:rdsR} - \eqref{eq:rdsE} is studied. For the general shape derived here, such a discussion can be carried out by performing suitable choices for functions $\Phi_i$, as e.g. those considered in \cite{wang2007classical}.}
			
			\section{Pattern formation analysis}
			\label{Sec4}
			
			The configuration of areas affected by damaged oligodendrocytes may be suitably investigated by means of a Turing instability analysis \cite{turing1990chemical} of the reaction-diffusion system with chemotaxis motion \eqref{eq:rdsR}-\eqref{eq:rdsE}. Furthermore, an investigation of diverse shapes of patterning, as well as their stability, can be obtained only through a deeper analysis of the problem, performing a higher-order expansion of the system.
			
			Let us set the problem by adding to system \eqref{eq:rdsR}-\eqref{eq:rdsE}  non-negative initial data 
			\begin{equation}\label{InCond}\nn
				\mathbf{W}(0,\bx)=\mathbf{W}_0(\bx)\geq 0, \mbox{ with} \mathbf{W}(t, \bx) = (M, C, D).
			\end{equation}
			and by imposing zero-flux conditions at the boundary, 
			\be\label{NoFlux}\nn
			\Big(\Phi_0(M)\,\nabla_{\bx}\, M-{\xi} \,\Phi_1(M) M\,\nabla_{\bx}\,C \Big)\cdot {\bf \widehat n}=0,\quad \nabla_{\bx} C\cdot {\bf \widehat n}=0,
			\ee
			being  ${\bf \widehat n}$ the external unit normal to {the boundary} $\pa\Gamma_{\bx}$.
			
			Patterns resulting from Turing instability emerge when an initially uniform and stable equilibrium becomes unstable because of the introduction of diffusive elements. Equating the right-hand side of \eqref{eq:rdsR}-\eqref{eq:rdsE} to zero, we can infer the existence of a microglia-free line $(0,\delta\,D,D)$ of unstable steady states, and a coexistence equilibrium $(M^*,C^*,D^*)=(1,\beta+\delta,1)$, which is always stable. 
			{
				
				The analysis of the conditions on parameters leading to the emergence of spatial patterns from a perturbation of equilibrium $(M^*,C^*,D^*)$ has been extensively carried out in the previously mentioned works \cite{lombardo2017demyelination,bilotta2018eckhaus,bisi2024chemotaxis,barresi2016wavefront} for a particular choice of involved functions. More specifically, a weakly nonlinear analysis of the problem has been carried out, including also wavefront invasion results. On the other hand, the analysis for the amplitude of the emerging pattern and the simulations proposed have been presented only for a one-dimensional setting or, in the case of Balo's Sclerosis \cite{lombardo2017demyelination}, in two dimensions with radial symmetry. Here we propose a two-dimensional weakly nonlinear analysis of the problem that allows us to investigate a richer pattern formation scenario  for the more general reaction-diffusion equations \eqref{eq:rdsR}-\eqref{eq:rdsE},  {  by taking the chemotactic sensitivity coefficient $\xi$ as bifurcation parameter}.
				
				In order to derive the amplitude equations, we perform a Taylor expansion of system  \eqref{eq:rdsR}-\eqref{eq:rdsE} up to the third
				order around the equilibrium $(M^*,C^*,D^*)$, writing
				\begin{equation}\label{SistComp1}
					\begin{aligned}
						\frac{\partial {\bf U}}{\partial t} &= \mathcal L\,{\bf U} + \mathcal H[{\bf U}], \quad \mbox{for} \quad
						{\bf U}=\left(\begin{array}{c}U\\V\\W\end{array}\right)=\left(\begin{array}{c}M- M^*\\C-C^*\\D-D^*\end{array}\right),
					\end{aligned}
				\end{equation}
				 { with the linear operator provided by $\mathcal{L}= \mathbb{A} + \mathbb{D} \Delta_{\bf x}$, being $\mathbb{A}$ and $\mathbb{D}$ the Jacobian and diffusion matrix, respectively, reported in \ref{Appendix}, as well as the remainder operator $\mathcal H[{\bf U}]$.}

				We just report here the necessary conditions for the formation of spatial patterns on the { chemotactic sensitivity coefficient $\xi$} \cite{turing1990chemical,murray2003ii}, { i.e. there exist a critical value
							\begin{equation}
					\xi_c=\frac{\left(\sqrt{-\theta\, \Pi'(M^*)} + \sqrt{\Phi_{0}(M^*)}\right)^2}{\Phi_{1}(M^*)}\label{ExpXic}, 
				\end{equation}
			 and the critical wavenumber
				\begin{equation}
				k_c^2=\sqrt{\frac{-\Pi'(M^*)}{\theta\, \Phi_{0}(M^*)}},
				\end{equation}}
				such that $\det({\mathbb A}-k_c^2{\mathbb D})=0$ and $\det({\mathbb A}-k^2{\mathbb D})>0$ for some wavenumber $k$ when $\xi>\xi_c$.}
		
			At this point, a deeper analysis of the model can provide more information on the shape and the stability of patterns { in a two-dimensional domain}, allowing for a better correlation with the real phenomenon. To this aim, we exploit the fact that, when the parameter  $\xi$ is close to the critical threshold, the change in the dynamics is slower. This allows us to investigate the formation of patterns employing amplitude equations.  {  More specifically, as illustrated e.g. in \cite{walgraef2012spatio}, each possible steady state of the reaction-diffusion dynamics considered corresponds to a planform characterized by $m$	pairs of wave vectors $({\bf k}_j , -{\bf k}_j )$ and, for critical wave vectors ($|{\bf k}_j|=k_c$), one has
			\be\label{SolAj}
			{\bf U}=\sum_{j=1}^m\left[{\bf A}_j(t)\,e^{i\,{\bf k_{\it j}\cdot\bx}}+\overline{{\bf A}}_j(t)\,e^{-i\,{\bf k_{\it j}\cdot\bx}}\right]
			\ee
			where ${\bf A}_j$ is the amplitude vector associated to the mode ${\bf k}_j$, and $\overline{{\bf A}}_j$ its complex conjugate.
			Depending on the value of $m$ and on the relation among wave vectors, one can analyze different patterns. In this work, we consider the case of $m=3$, i.e. we express the solution by means of three active dominant resonant pairs of eigenmodes ${\bf k}_j$, $j=1,2,3$, individuating angles of $2\pi/3$, with $|{\bf k}_j| = k_c$ and such that ${\bf k}_1 + {\bf k}_2 + {\bf k}_3 = {\bf 0}$. This is one of the classical choices relevant to pattern formation theory in two dimensions \cite{walgraef2012spatio,rucklidge2009design}. Geometrically, this enforces a mutual $120^{\circ}$ separation in Fourier space. 
			}
			
			{
				Around the bifurcation value, the formation and development of patterns occur when $\xi > \xi_c$. To analyze this scenario, we express the bifurcation parameter $\xi$ as follows
				\begin{equation}\label{expxi}
					\xi=  \xi_c +\eta\,\xi_1+\eta^2\,\xi_2+\eta^3\,\xi_3+O(\eta^4),
				\end{equation}
				where $\eta$ is a small parameter. Analogously, we expand the solution vector ${\bf U}$  in terms of $\eta$
				\begin{equation}\label{UExp}
					{\bf U} = \eta
					\left(
					\begin{array}{c}U_1 \\ V_1\\W_1 \end{array}
					\right)
					+ \eta^2 
					\left(
					\begin{array}{c}U_2 \\ V_2\\W_2 \end{array}
					\right)
					+ \eta^3
					\left(
					\begin{array}{c}U_3 \\ V_3\\W_3 \end{array}
					\right)
					+ O(\eta^4).
				\end{equation}
				When the bifurcation parameter is close to the threshold, the pattern's amplitude undergoes slow temporal evolution. Consequently,  {  we adopt the multiple time scales method, for which the time derivative can be expanded as 
				\begin{equation}\label{timeExp}
					\frac{\partial}{\partial t} = \eta \frac{\partial}{\partial T_1} + \eta^2 \frac{\partial}{\partial T_2} + O(\eta^3),
				\end{equation}
				where $T_1=\eta\,t$ and $T_2=\eta^2 t$ allow us to distinguish between the fast and slow time scales and to avoid secular terms that may grow boundlessly \cite{kevorkian2012multiple}.}
				{
					We underline here that, differently from previous approaches where the scale $T_1$ is neglected, we include it, allowing for an expansion of the amplitude of patterns themselves.}		
				 {  We recall here that time variable is already scaled of order $\varepsilon$ with respect to the original scale, after the parabolic limit, which means that the multiple time scale is actually $ \varepsilon\eta \dfrac{\partial}{\partial T_1} + \varepsilon \eta^2 \dfrac{\partial}{\partial T_2} + O(\varepsilon\,\eta^3)$.}
				 {
By substituting the expansions \eqref{UExp}-\eqref{timeExp} in system \eqref{SistComp1} and collecting terms at the same order of $\eta$, the terms of the expansion \eqref{UExp} corresponding to the first two orders result of the form 
\begin{equation}\label{Expu1Text}
	\left(
	\begin{array}{c}U_1\\V_1\\W_1 \end{array}
	\right)=\left(
	\begin{array}{c}\rho\\1\\0 \end{array}
	\right)\sum_{j=1}^3\left[{\mathcal W}_j(t)\,e^{i\,{\bf k_{\it j}\cdot\bx}}+\overline{\mathcal W}_j(t)\,e^{-i\,{\bf k_{\it j}\cdot\bx}}\right],
\end{equation}
with $\overline{\mathcal W}_j$ denoting the complex conjugate and
$\rho=1 + \sqrt{\dfrac{-\theta\,\Pi'(M^*)}{\Phi_0(M^*)}}$,
and
\begin{equation}\label{Expu2Text}
	\begin{aligned}
		\left(
		\begin{array}{c}U_2 \\ V_2\\W_2 \end{array}
		\right)
		=& \left(
		\begin{array}{c}X_0 \\ Y_0\\Z_0 \end{array}
		\right)\left(|\mathcal W_1(t)|^2+|\mathcal W_2(t)|^2+|\mathcal W_3(t)|^2\right)+\sum_{j=1}^3 \left(
		\begin{array}{c}\rho \\ 1\\0 \end{array}
		\right)\,\mathcal V_j(t)\,e^{i\,\bk_j\cdot\bx}\\
		&+\sum_{j=1}^3 \left(
		\begin{array}{c}X_{2} \\ Y_{2}\\Z_{2}\end{array}
		\right)\,\mathcal W_j^2(t)\,e^{2\,i\,\bk_j\cdot\bx}+\sum_{\substack{j=1,2,3 \\ l\equiv j+1\mbox{ \tiny mod}3}} \left(
		\begin{array}{c}X_{1} \\ Y_{1}\\Z_{1}\end{array}
		\right)\,\mathcal W_j(t)\,\overline{\mathcal W}_l(t)\,e^{\,i\,(\bk_j-\bk_l)\cdot\bx}+\mbox{ c.c.}
	\end{aligned}
\end{equation}
where c.c. denotes the complex conjugate.
The explicit computation for expressions \eqref{Expu1Text}-\eqref{Expu2Text} and coefficients in \eqref{Expu2Text} are provided in \ref{Appendix}.}

				By  putting together \eqref{Expu1Text}, and \eqref{Expu2Text}, we obtain the expression for the amplitude { ${\bf A}_j=(A_{j}^U,A_{j}^V,A_{j}^W)^T$  appearing in (\ref{SolAj})} in expanded form as}
				\begin{equation}\label{exprAj}
				{\bf A}_j=\eta\, \left(
				\begin{array}{c}\rho \\ 1\\0 \end{array}
				\right)\,\mathcal W_j+\eta^2\, \left(
				\begin{array}{c}\rho \\ 1\\0 \end{array}
				\right)\,\mathcal V_j+O(\eta^3),\quad j=1,2,3.
				\end{equation}
					{Then, the equations for amplitudes read as
					\begin{equation}\label{EqAj}
						\frac{\pa {\bf A}_j}{\pa t}=\eta^2\, \left(
						\begin{array}{c}\rho \\ 1\\0 \end{array}
						\right)\,\frac{\pa\mathcal W_j}{{\pa T_1}}+\eta^3\, \left(
						\begin{array}{c}\rho \\ 1\\0 \end{array}
						\right)\,\left(\frac{\pa\mathcal W_j}{{\pa T_2}}+\frac{\pa\mathcal V_j}{{\pa T_1}}\right)+O(\eta^4),\quad j=1,2,3.
				\end{equation}}
		{{ The terms containing the derivatives with respect to $T_1$ and $T_2$ are recovered in \ref{Appendix}, and allow us to write the evolution equation for the amplitudes $A_{j}^U$ as follows}
		\begin{equation}\label{SistAj}
			r_0\frac{\pa A_j^U}{\pa t}=\xi_m \,A_j^U+\left(s_1+\xi_m\,\tilde s_1\right)\,\overline {A_l^U}\,\overline{A_m^U}+A_j^U\,\left[s_2\,|{A_j^U}|^2+s_3\left(|{A_l^U}|^2+|A_m^U|^2\right)\right].
		\end{equation}	
		 An analogous expression may be obtained for $A_j^V$, while for the variable $W$ we do not have the evolution of the corresponding amplitude, since this variable is not affected by diffusion processes.
		 { The term $\xi_m=\dfrac{\xi-\xi_c}{\xi_c}$ represents the magnitude of the perturbation, while the remaining coefficients in \eqref{SistAj} are the outcome of the computations detailed in the \ref{Appendix}.}
		{We decompose each amplitude into its mode and phase angle, that is  $A_j^U=\rho_j\,e^{i\,\phi_j}$ and, by splitting the real and imaginary parts, we obtain the following system:
			\begin{equation}\label{SistRhoPhi}
				\begin{aligned}		
					r_0\, \frac{\partial \phi}{\partial t} &= -\left(s_1+\xi_m\,\tilde s_1\right)\frac{\rho_1^2\,\rho_2^2+\rho_2^2\,\rho_3^2+\rho_1^2\,\rho_3^2}{\rho_1\,\rho_2\,\rho_3}\,\sin(\phi) \\
					r_0\,\frac{\partial \rho_1}{\partial t} &= \xi_m\,\rho_1 +\left(s_1+\xi_m\,\tilde s_1\right) \, \rho_2 \, \rho_3 \, \cos(\phi) +s_2 \, \rho_1^3 +s_3\, (\rho_2^2 + \rho_3^2) \, \rho_1
					\\
					r_0\,\frac{\partial \rho_2}{\partial t} &= \xi_m \,\rho_2  +\left(s_1+\xi_m\,\tilde s_1\right) \, \rho_1 \, \rho_3 \, \cos(\phi) +s_2 \, \rho_2^3 +s_3\, (\rho_1^2 + \rho_3^2) \, \rho_2\\
					r_0\,\frac{\partial \rho_3}{\partial t} &=\xi_m \,\rho_3 +\left(s_1+\xi_m\,\tilde s_1\right) \, \rho_1 \, \rho_2 \, \cos(\phi) +s_2 \, \rho_3^3 +s_3\, (\rho_1^2 + \rho_2^2) \, \rho_3,
				\end{aligned}
			\end{equation}
			being $\phi=\phi_1+\phi_2+\phi_3$.
		}
		
		{Stationary states of system \eqref{SistRhoPhi} correspond to the different observable patterns. In particular, we can individuate the following ones:
			\begin{itemize}
				\item[i)] Homogeneous solution with $\rho_1=\rho_2=\rho_3=0$; in this case, no pattern emerges.
				\item[ii)] Equilibrium $\mathcal S=\left(\phi,\rho_1,\,0,\,0\right)$, with $\rho_1=\sqrt{-\dfrac{\xi_m}{s_2}}$.  {  In this case, the solution expressed in formula \eqref{SolAj} reduces to a single contribution given by a periodic function along the direction individuated by $\mathbf{k}_1$,
leading to striped patterns.}
				\item[iii)]  Equilibria $\mathcal H_{\bar\phi}^\pm=\left(\bar\phi,\,\bar\rho_{\pm},\,\bar\rho_{\pm},\,\bar\rho_{\pm}\right)$, with $$\bar\phi=\frac \pi2\left(1+\mbox{sign}\left(s_1+\xi_m\,\tilde s_1\right)\right),$$ $$\bar\rho_{\pm}=\frac{
					|s_1 + \xi_m\,\tilde s_1| \pm \sqrt{-4\,(s_2+2s_3)\,\xi_m + \left(s_1+\xi_m\,\tilde s_1\right)^2}}{2\,(s_2 + 2 s_3)}.
				$$
				 {  The solution given in formula \eqref{SolAj} is composed by three periodic functions having the same amplitude along the three directions individuated by $\mathbf{k}_j$,
					leading to hexagonal patterns.}
				\item[iv)]  Equilibria  $\mathcal M_{\tilde\phi}=\left(\tilde\phi,\,\tilde\rho_{1},\,\tilde\rho_{2},\,\tilde\rho_{2}\right)$, with  $$\tilde\phi=\frac \pi2\left(1-\mbox{sign}\left(\frac{s_1+\xi_m\,\tilde s_1}{s_2-s_3}\right)\right),$$ $$\tilde\rho_1=\left|\frac{s_1+\xi_m\,\tilde s_1}{s_2-s_3}\right|,\quad
				\tilde\rho_2=\sqrt{\frac{-\xi_m - s_2\, \tilde\rho_1^2}{s_2+s_3}}.
				$$  {  As in the previous case, the solution expressed in formula \eqref{SolAj} is composed by three contributions, but only two of the three periodic functions share the same amplitude,
					leading to mixed patterns (elongated hexagons).}
			\end{itemize}	
		}
		
		We observe that the existence and stability of patterns strongly depend on the sign of functions $\Phi_0$, $\Phi_1$, $\Pi$, and their derivatives. For this reason, we choose to investigate it numerically for specific expressions of them.
		
		\section{Numerical simulations}
		\label{sec5}
		
		In this section, we move from the general system \eqref{eq:rdsR}-\eqref{eq:rdsE} to a more specific formulation, to analyze numerically the formation of patterns and some stability results.
		
		We take inspiration from \cite{lombardo2017demyelination}, where the authors consider a constant diffusion rate for macrophages, a modified version of the Keller–
		Segel equations, which include a ``volume-filling" effect for the chemotactic term, and a logistic term to describe the proliferation and saturation of microglia.
		This model can be recovered  starting from our kinetic description by setting functions and parameters in such a way we get, at macroscopic level, 
		\be
		\Phi_0(M)\equiv 1,\quad\Phi_1(M)=\dfrac{M}{M+1},\quad\Pi(M)=M\,(1-M). \label{FunsSim}\ee
		The model proposed in \cite{lombardo2017demyelination}
		has been investigated providing conditions for pattern formation and performing weakly nonlinear analysis in one dimension. Here we apply the procedure described in the previous section to obtain two-dimensional (and hence richer) depictions.
		
		First of all, we set the parameters of the model as done in \cite{lombardo2017demyelination}, i.e.
		\begin{equation}
			\tau=1,\quad\beta=1,\quad\delta=1,\quad r=1,\quad \nu=1.\label{ParsSim}
		\end{equation}
		By computing all the coefficients appearing in the system \eqref{SistRhoPhi}, we may obtain results on the existence and stability of striped, hexagonal, or mixed patterns for varying values of the cytokines diffusion coefficient $\theta$ and the normalized distance of the chemotactic rate $\xi$ from the critical value $\xi_c$, i.e. the quantity $\xi_m$ defined in \eqref{CoefAmp}; the complete scenario is depicted in Figure \ref{fig1}.
		\begin{figure}[ht!]
			\centering
			\includegraphics[scale=0.45]{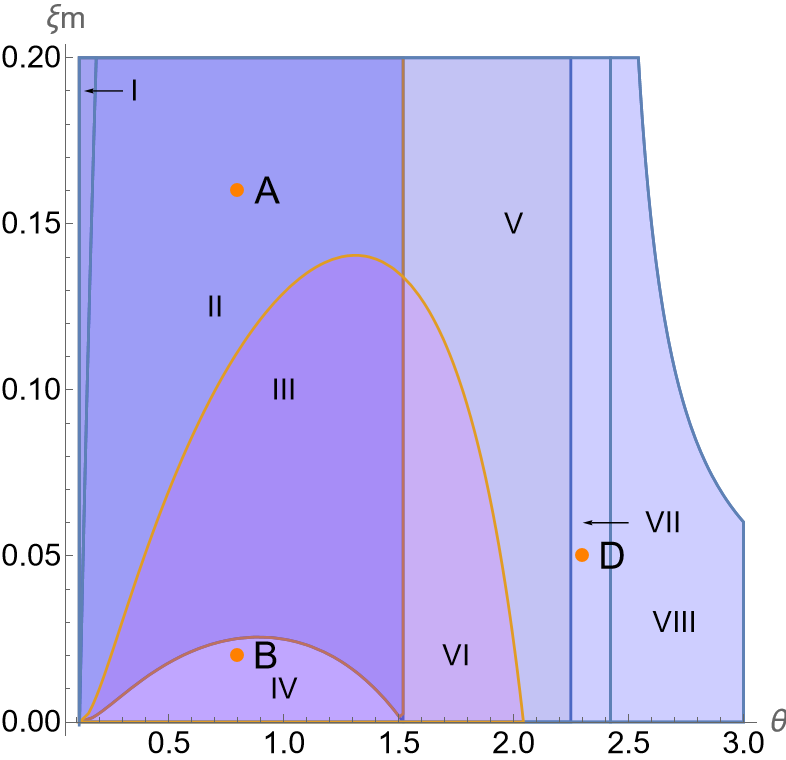}
			\includegraphics[scale=0.45]{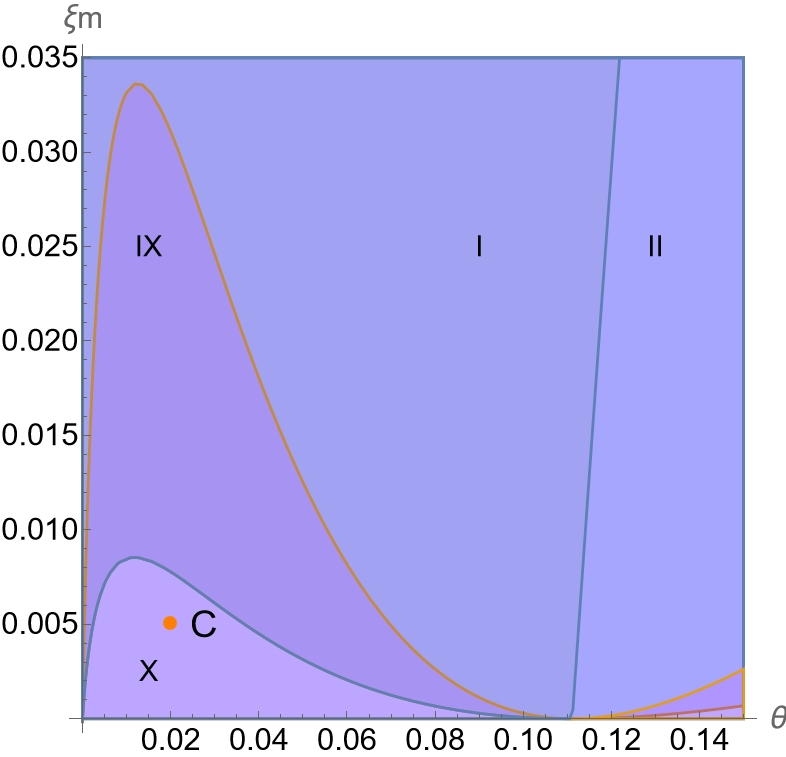}
			\caption{Regions of existence and stability of steady states of system \eqref{SistRhoPhi} for varying parameters $\theta$ and $\xi_m$, choosing functions as in \eqref{FunsSim} and parameters as in \eqref{ParsSim}, with a zoom of the area near the point $(0,0)$ on the right panel.  { Labeled points in the plane correspond to the values taken for numerical simulations.}}
			\label{fig1}
		\end{figure}
		Although the range for realistic values of the cytokines diffusion coefficient $\theta$ reported in the literature \cite{lombardo2017demyelination} is between $0.5$ and $1.5$, we propose here a complete analysis, also outside this range, in view of a future comparison with other models, such as the modified one involving Allee effect \cite{bisi2024chemotaxis}. For this reason, a magnification of the graph in the left panel of Figure \ref{fig1} around the vertex $(0,0)$ is reported on the right side. 
		
		The conditions for the existence and stability of equilibria discussed above lead to a partition of the space of parameters in several regions, labeled by roman numbers in Figure \ref{fig1}; for each region, the admissible equilibria are individuated and listed in Table \ref{tab1}, where the stable ones are highlighted in bold and red.
		\begin{table}[]
			\centering
			 { 
				\caption{Existence and stability of steady states of system \eqref{SistRhoPhi} (stripes ${\mathcal S}$, hexagons $\mathcal{H}_l^\pm $, mixed $\mathcal{M}_l, l=0,\pi$) in each region of Figure \ref{fig1}, choosing functions as in \eqref{FunsSim} and parameters as in \eqref{ParsSim}. Stable equilibria are highlighted in bold and red.}\label{tab1}}
			\begin{tabular}{|c|c|}
				\hline
				Area & Equilibria\\
				\hline
				I & $ {\color{red}{\boldsymbol{{\mathcal S}}}},\,\mathcal{H}_0^-,\,\mathcal{H}_\pi^-,\,\mathcal{M}_0$\\
				\hline
				II & $ {\color{red}{\boldsymbol{{\mathcal S}}}},\,\mathcal{H}_0^-,\,\mathcal{H}_\pi^-,\,\mathcal{M}_\pi$\\
				\hline
				III & $ {\color{red}{\boldsymbol{{\mathcal S}}}},\,\mathcal{H}_0^-,\, {\color{red}{\boldsymbol{{\mathcal H_\pi^-}}}},\,\mathcal{M}_\pi$\\
				\hline
				IV & ${{{{\mathcal S}}}},\,\mathcal{H}_0^-,\, {\color{red}{\boldsymbol{{\mathcal H_\pi^-}}}},\,\mathcal{M}_\pi$\\
				\hline
				V & $\mathcal{H}_0^-,\,{{{{\mathcal H_\pi^-}}}},\,\mathcal{M}_\pi$\\
				\hline
				VI & $\mathcal{H}_0^-,\, {\color{red}{\boldsymbol{{\mathcal H_\pi^-}}}},\,\mathcal{M}_\pi$\\
				\hline
				VII & $\mathcal{H}_0^-,\,{{{{\mathcal H_\pi^-}}}}$\\
				\hline
				VIII & $\mathcal{H}_0^+,\,{{{{\mathcal H_0^-}}}}$\\
				\hline
				IX & $ {\color{red}{\boldsymbol{{\mathcal S}}}},\, {\color{red}{\boldsymbol{{\mathcal H_0^-}}}},\,\mathcal{H}_\pi^-,\,\mathcal{M}_\pi$\\
				\hline
				X & ${{{{\mathcal S}}}},\, {\color{red}{\boldsymbol{{\mathcal H_0^-}}}},\,\mathcal{H}_\pi^-,\,\mathcal{M}_\pi$\\
				\hline     
			\end{tabular}
		\end{table}
		
		Consequently, we pick values for $\theta$ and $\xi_m$ in different regions and perform numerical simulations using the online software VisualPDE \cite{walker2023visualpde} in a square domain of size $6\,\pi$, starting from a random perturbation of equilibrium $\left(M^*,\, C^*,\, D^*\right)$, { and by imposing no-flux boundary conditions}. In particular, we show the formation of patterns for the microglia population $M$. 
		
		We start with region II, taking $\theta=0.8$ (that provides $\xi_c\approx 7.18$) and $\xi_m=0.16$  { (point A)} (corresponding to $\xi\approx 8.33$ by means of \eqref{ExpXic} and \eqref{CoefAmp}). According with Table \ref{tab1}, we have the stability of the striped pattern, and this can be observed also numerically in Figure \ref{fig2}, Panel (a). 
		\begin{figure}[ht!]
			\centering
			
			\begin{tabular}{c c c}
				(a)&(b)&(c)\\
				\includegraphics[scale=0.33]{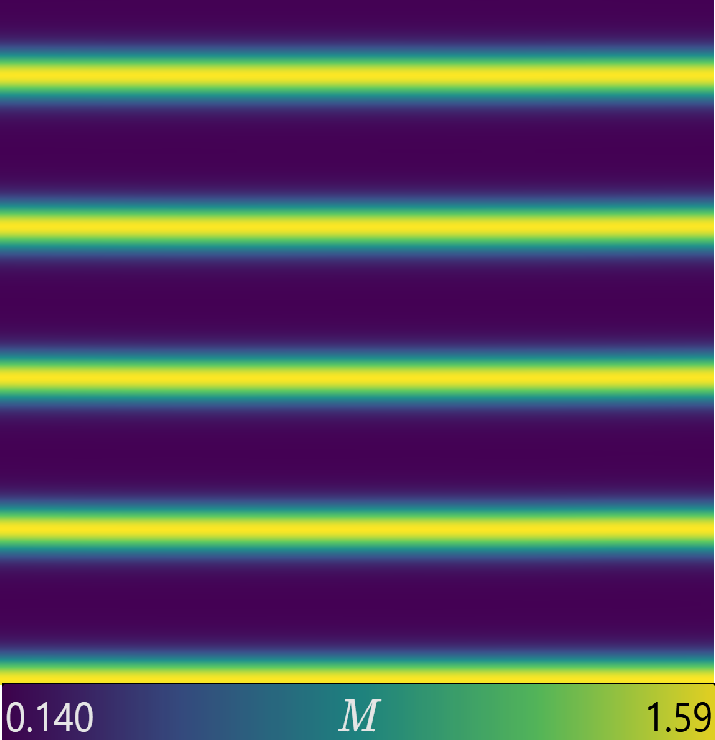} 
				&
				\includegraphics[scale=0.33]{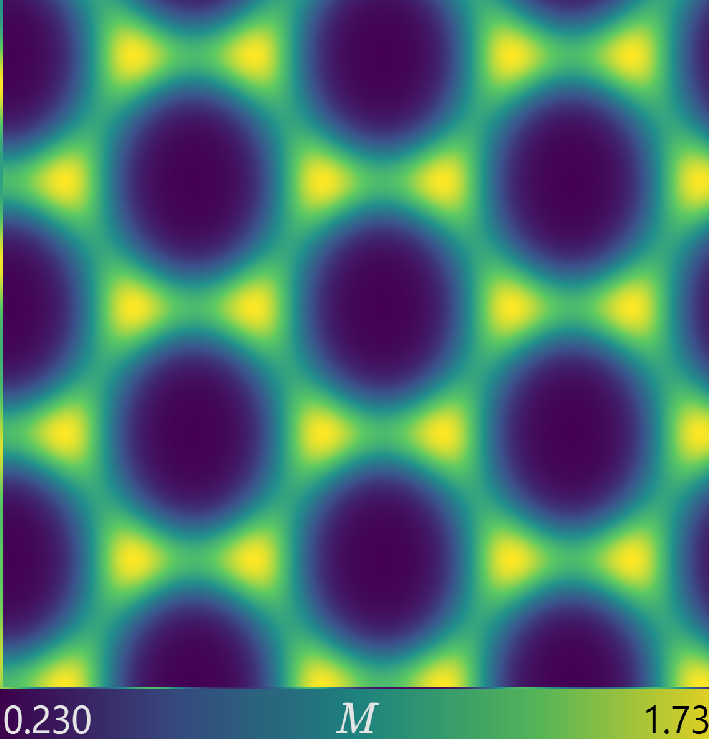}
				&
				\includegraphics[scale=0.33]{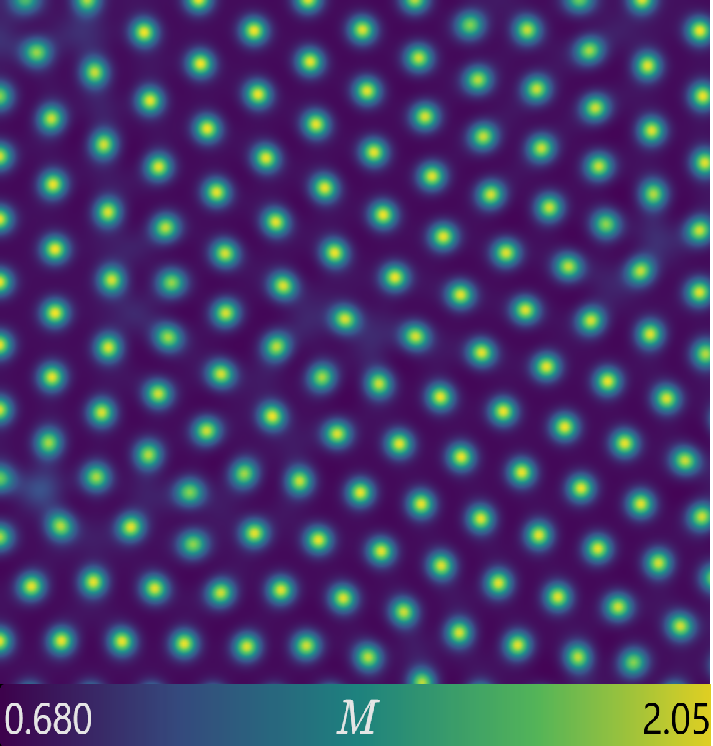}
			\end{tabular}
			\caption{Long-time patterning of microglia population, described by system \eqref{eq:rdsR}-\eqref{eq:rdsE}, taking functions as in \eqref{FunsSim}, parameters as in \eqref{ParsSim}. Panel (a): $\theta=0.8$ and $\xi\approx 8.33$ { (point A in Figure \ref{fig1})}.
			Panel (b): $\theta=0.8$ and $\xi\approx 7.32$
			{ (point B in Figure \ref{fig1})}. Panel (c): $\theta=0.02$ and $\xi\approx 2.62$
			{ (point C in Figure \ref{fig1})}.}
			\label{fig2}
		\end{figure}
		If $\xi_m$ decreases to $0.02$ (corresponding to $\xi\approx 7.32$), we move to region IV  { (point B)}; as expected, we get a hexagonal stable pattern, as shown in Figure \ref{fig2}, Panel (b).
		{ We observe a similar scenario, by considering values in region X ($\theta=0.02$,  $\xi_c\approx 2.6$, $\xi_m=0.005$,  { point C}, and hence $\xi\approx 2.62$).} However, since both cytokines diffusion coefficient and chemotactic sensitivity are lower, the microglia population tends to cluster in spots, as can be seen in Figure \ref{fig2}, Panel (c).
		On the contrary, higher values for $\theta$ and $\xi_m$ ($\theta=2.2$, $\xi_c\approx 12.34$, $\xi_m=0.05$ and $\xi\approx 13.0$), corresponding to region V  { (point D)}, induce a different scenario characterized by unstable solutions, oscillating between the two hexagonal and the mixed pattern. Figure \ref{fig6} reports this behavior at four different time values. { We have also checked that values in region VII lead to oscillating patterns between the two hexagonal types, while for values in region VIII no pattern arises.}
		\begin{figure}[ht!]
			\centering
				\begin{tabular}{c c}
				(a)&(b)\\
			\includegraphics[scale=0.33]{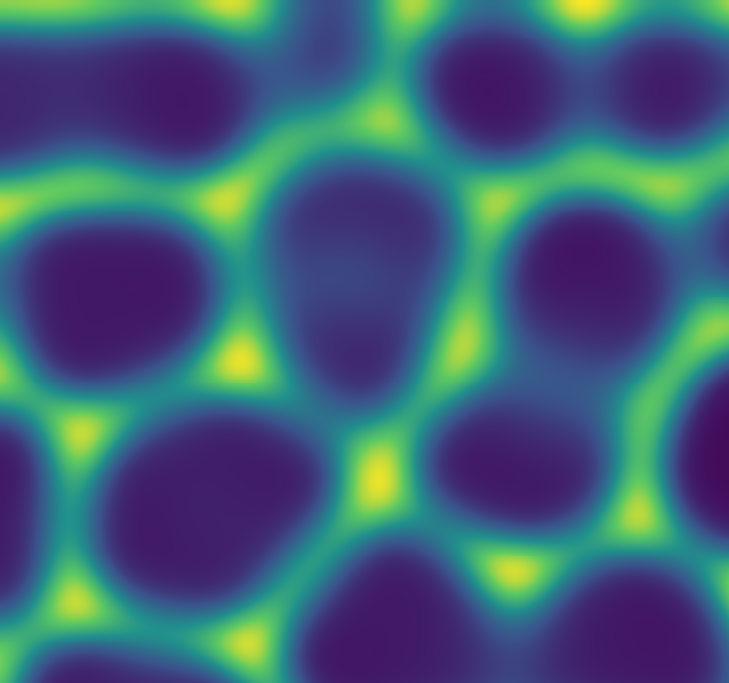}&
			\includegraphics[scale=0.33]{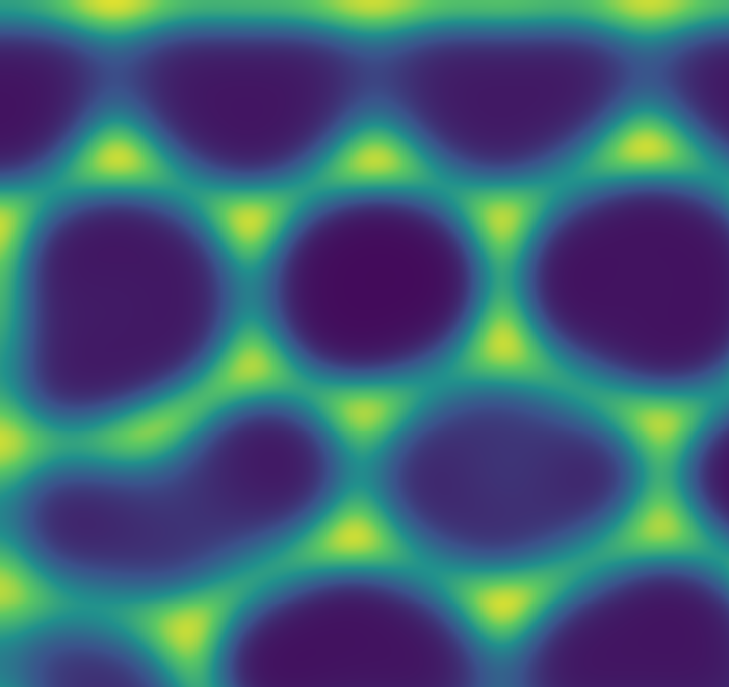}\\
				(c)&(d)\\
			\includegraphics[scale=0.33]{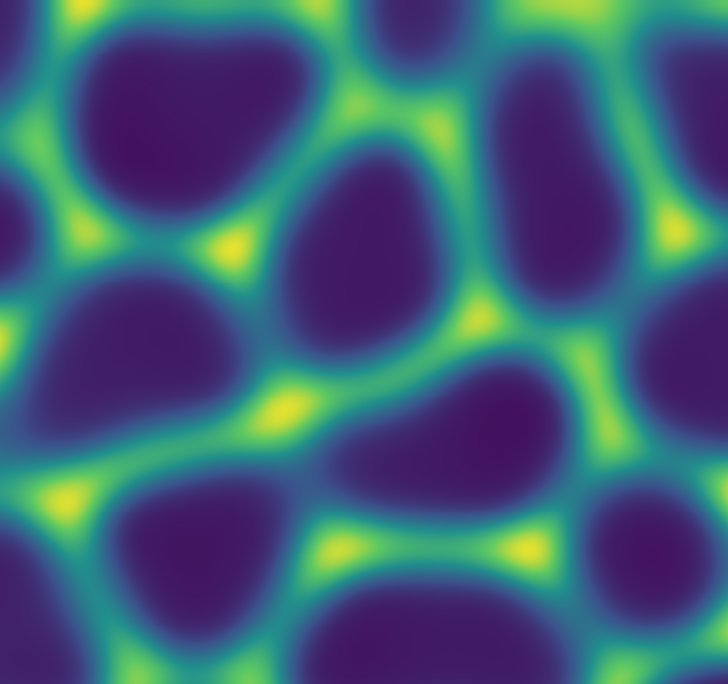}&
			\includegraphics[scale=0.33]{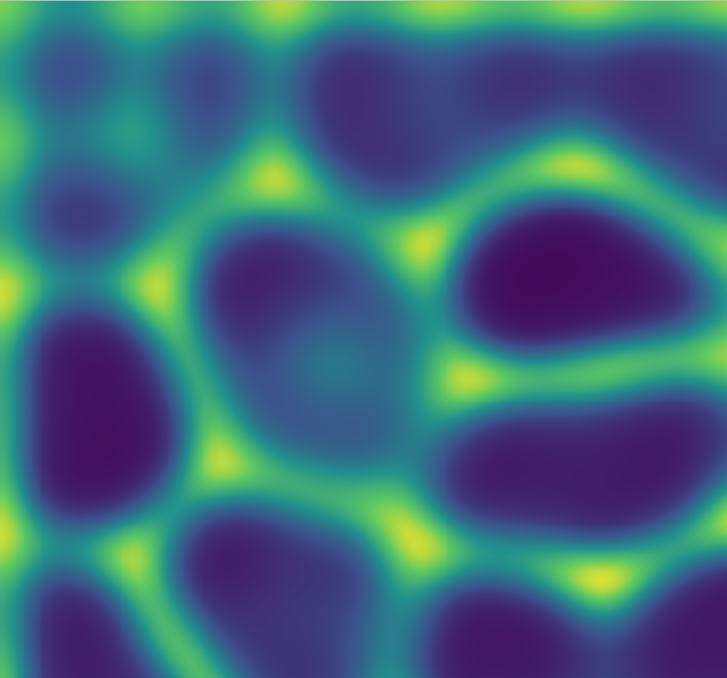}\\
			\end{tabular}\\
			\includegraphics[scale=0.27]{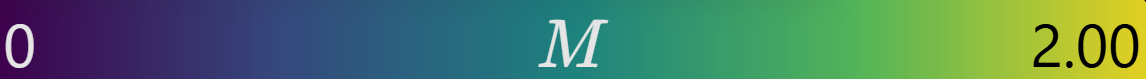}
			\caption{Oscillating in time of patterning for microglia population, described by system \eqref{eq:rdsR}-\eqref{eq:rdsE}, taking functions as in \eqref{FunsSim}, parameters as in \eqref{ParsSim}, $\theta=2.2$ and $\xi\approx 13$ { (point D in Figure \ref{fig1})}, at time $t=700$ { in Panel (a)}, $t=800$ { in Panel (b)}, $t=900$ { in Panel (c)}, and $t=1000$ { in Panel (d)}.}
			\label{fig6}
		\end{figure}
		
		{ As an additional case, we simulate the dynamics of the complete macroscopic system \eqref{MacComp1}–\eqref{MacComp5}, starting from a specific initial condition: in the spatial domain, microglia $M$ are set to zero everywhere except at a single spot; $A$ is uniformly equal to one; $S$ is defined in accordance with \eqref{MacComp2}; and both $C$ and $D$ are identically zero throughout the domain. We aim at describing the evolution in time of all the quantities involved in the model, thus we fix parameters as above and consider the case $\theta=0.02$ and $\xi\approx 2.62$, leading to hexagonal patterns, and $\Theta=0.5$. The outcome of the simulation performed on a square domain of size $2\,\pi$ is reported in Figure \ref{figall}. In particular, we show results at time $t=0$ (first column), $t=6$ (second column), $t=16$ (third column), $t=700$ (fourth column). We observe that at the first stage quantities $S$ (first row) and $M$ (sond row) start to diffuse, while for $C$ (third row) and $D$ (fourth row) we have an initial production. Then, all the quantities reach a configuration which is closer to the homogeneous equilibrium and, after longer time, we have the formation of patterns for $S$, $M$, and $C$, while the oligodendrocytes $D$ are totally consumed. We omit to report here the behavior of $A$ since it is constant in time and space.
		
		\begin{figure}[ht!]
			\centering
			\begin{tabular}{c}
				(a)\\
				\includegraphics[scale=0.28]{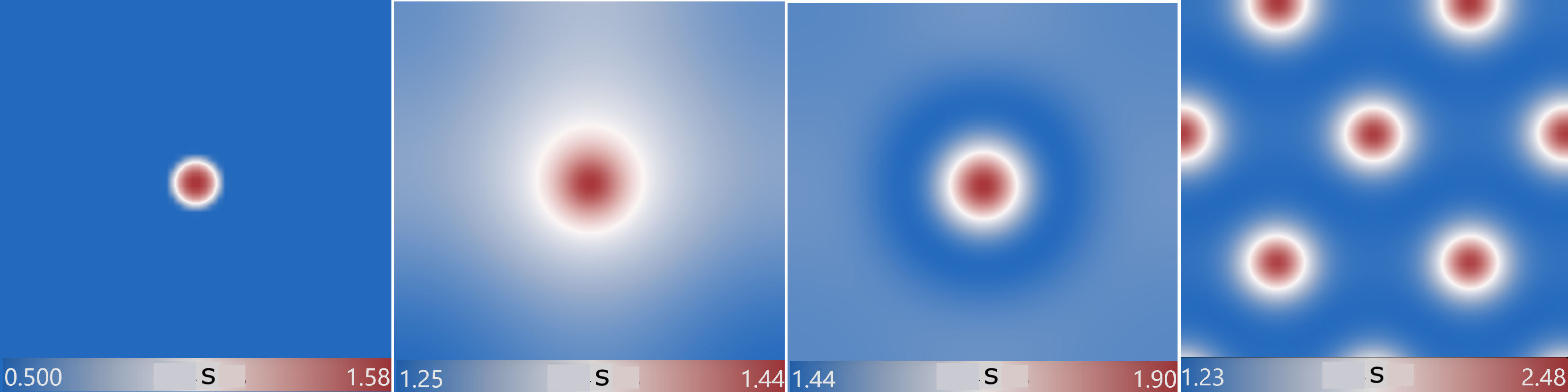}\\
				(b)\\
				\includegraphics[scale=0.28]{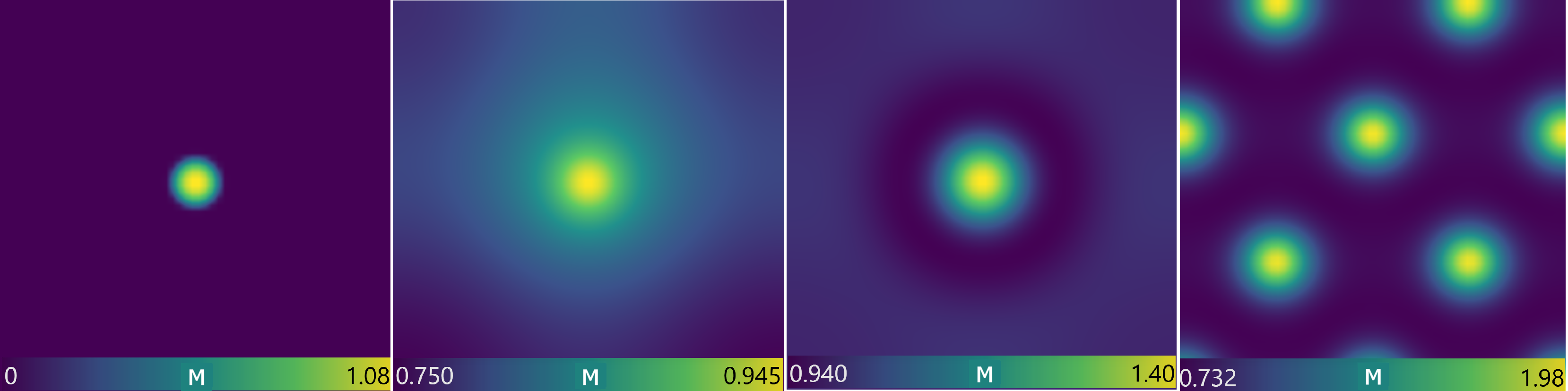}\\
				(c)\\
				\includegraphics[scale=0.28]{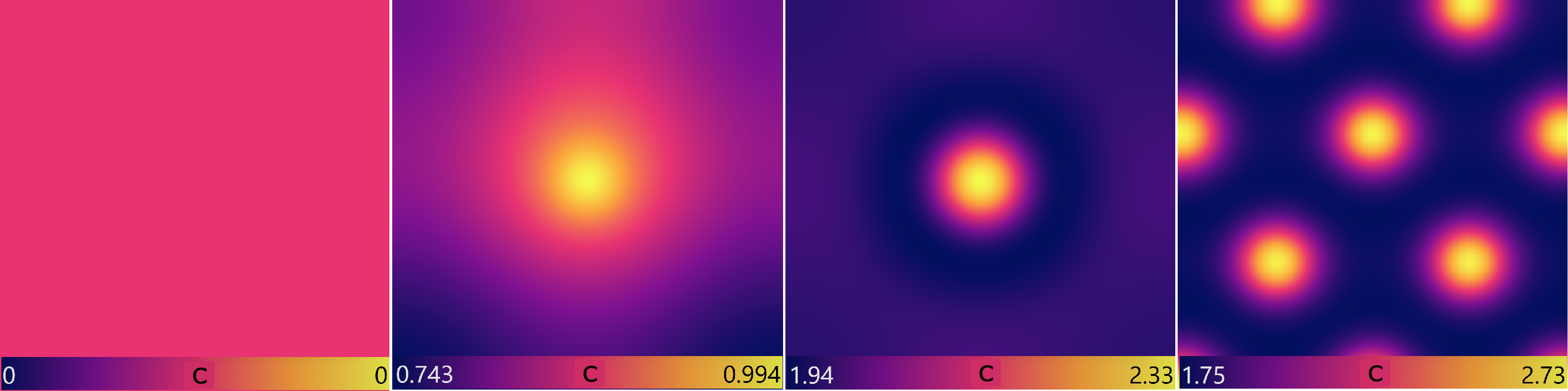}\\
				(d)\\
				\includegraphics[scale=0.28]{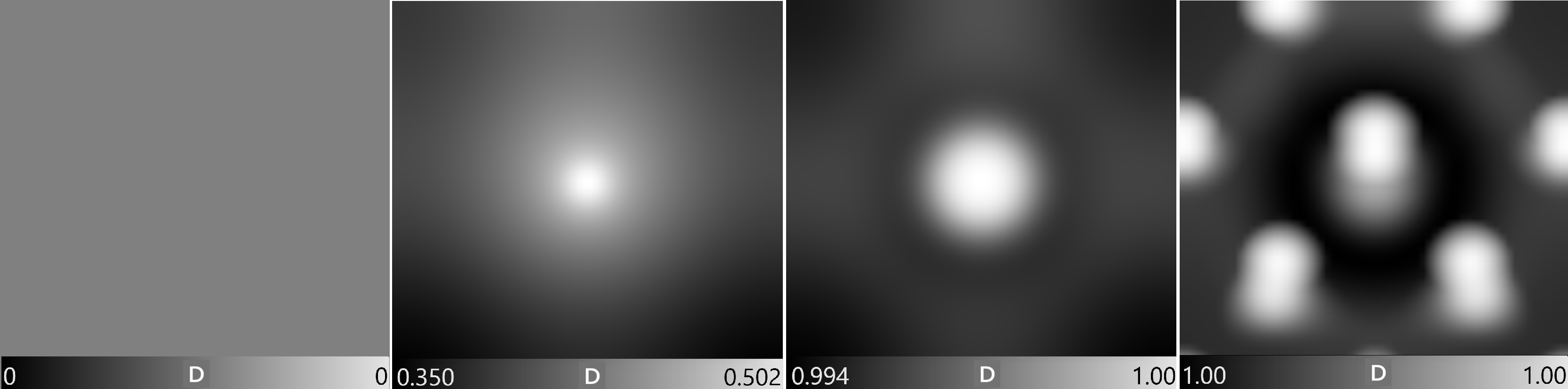}\\
			\end{tabular}
			\caption{{ Behavior in time of immunosuppressive cells $S$, in Panel (a), microglia $M$, in Panel (b), cytokines $C$, in Panel (c), destroyed oligodendrocytes $D$, in Panel (d), described by system \eqref{MacComp1}-\eqref{MacComp5}, taking functions as in \eqref{FunsSim}, parameters as in \eqref{ParsSim}, $\Theta=0.5$, $\theta=0.02$ and $\xi\approx 2.62$, at time $t=0$ (first column), $t=6$ (second column), $t=16$ (third column), $t=700$ (fourth column).}}
			\label{figall}
		\end{figure}

		}
		
		\section{Concluding remarks and perspectives}
		\label{sec6}
		
		In this paper, we have derived a class of models, which can  {  describe the cellular mechanisms behind the} emergence of type III lesions due to Multiple Sclerosis. The reaction-diffusion equations modeling the population dynamics at the macroscopic level have been obtained as a diffusive limit of a proper mesoscopic description, based on the kinetic theory of active particles. This derivation has the great advantage of relating the macroscopic dynamics with the microscopic interactions; more precisely, the macroscopic parameters, usually derived from experimental observations and heuristic considerations, can be set properly, in accordance with the underlying microscopic mechanism.
		The resulting model has been studied to investigate the formation of patterns. The Turing instability analysis, providing only necessary conditions for the emergence of spatially periodic solutions  {  for cell compartments}, has been integrated with weakly nonlinear analysis, allowing the prediction of the shape and stability of patterns. Such analysis, performed in two-dimensional domains, extends previous results in 1-D case \cite{lombardo2017demyelination,bisi2024chemotaxis}. Some simulations have been performed to validate the theoretical results and extend the discussion far from the bifurcation value, where the weakly nonlinear analysis fails.  {  The numerics confirm in 2-D case a rich scenario, where
		 spatial striped and hexagonal patterns  {  for microglia and cytokines can emerge for varying parameters.
		We emphasize that the present analysis aims to describe the cellular dynamics leading to type III plaque formation, without attempting to reproduce the lesions typically reported in medical literature. Unlike in \cite{oliveira2024reaction,travaglini25}, our model does not include a variable for consumed myelin. Instead, it focuses solely on the lysis of oligodendrocytes, which is the process responsible for myelin loss. Nevertheless,}
		  these results may lead to a better understanding of activation mechanisms leading to different shapes of plaques as those reported in literature for atypical demyelinating lesions \cite{filippi2019assessment,Litsou25}.}
		In addition, far from the critical value of the bifurcation parameter, it can be observed the formation of oscillating patterns, whose shape changes over time.
		 {  Also in this case, this result offers a good starting point toward the investigation of evolution of lesions, as observed in the medical literature \cite{calabrese2008morphology}.} 
		
		The analysis of the pattern formation has been proposed for the more general formulation of the model but it has been discussed numerically only for a specific choice of parameters and functions in diffusive and growth processes. As future work, it would be of great interest to analyze different mechanisms in diffusive and chemotactic terms, as done in \cite{bisi2024chemotaxis}, where the logistic growth for microglia has been compared with a cubic function taking into account the Allee effect
		 {  and in \cite{gargano2024cytokine}, where the influence of cytokines on macrophages activation is included. Additionally, the structure of our model allows for the incorporation of additional cell types and molecular mechanisms characteristic of distinct MS lesion types, { as performed in \cite{moise2021mathematical}}. Type III lesions, indeed, are characterized by 
			predominance of microglia and oligodendrocyte dysfunction. In contrast, Type II lesions display 
			significant T cell infiltration, active myelin degradation, and remyelination processes.		
		By integrating these specific cellular and biochemical dynamics, the model has the potential to show under which conditions patterns observed in Type III or II arise. { More specifically, in this model, the variables typically involved in Type II lesions already considered in \cite{oliveira2024reaction}, that are self-reactive T-cells migrating into the central nervous system and the myelin sheath attacked by these ones or restored by oligodendrocytes, can be included in the two-dimensional modeling. This would provide more clinically relevant insights into lesion heterogeneity and progression.}
		
		Moreover, given the dynamic and irregular geometry of active MS plaques, as future development the domain evolution can be incorporated into the model, by using tools for pattern formation on growing or deforming domains, such as those in \cite{van2021turing}. Finally, recent studies \cite{attfield2022immunology} have reconsidered the notion of an exclusively antigen-specific cause of MS, showing that different immune cell types can share common functions that contribute to disease progression. They also emphasize the importance of environmental context in shaping immune cell phenotypes and reveal that the pathogenic differentiation of these cells may be reversible through therapeutic intervention. Including dynamic immune cell states and reversible pathogenic phenotypes in a future work could better capture MS mechanisms and treatment responses.}

		\vspace{1cm}
		
		\emph {Acknowledgments} 
		This work was performed in the frame of activities sponsored by the Italian National Group of
		Mathematical Physics (GNFM-INdAM) and by the University of Parma (Italy) and Pavia (Italy). 
		The authors MB, MG, GM thank the support of the project PRIN 2022 PNRR "Mathematical
		Modelling for a Sustainable Circular Economy in Ecosystems" (project code P2022PSMT7, CUP D53D23018960001) funded by the European Union - NextGenerationEU PNRR-M4C2-I 1.1 and by MUR-Italian Ministry of Universities and Research. The authors MB, MG and RT also thank the support of the University of Parma through the action Bando di Ateneo 2022 per la ricerca, cofunded by MUR-Italian Ministry of Universities and Research - D.M. 737/2021 - PNR - PNRR - NextGenerationEU (project "Collective and Self-Organised Dynamics: Kinetic and Network Approaches").
		RT is a post-doc fellow supported by the National Institute of Advanced Mathematics (INdAM), Italy. 
		The work of RT  was carried out in the frame of activities sponsored by the Cost Action CA18232, by the Portuguese Projects UIDB/00013/2020 (\url{https://doi.org/10.54499/UIDB/00013/2020}), UIDP/00013/2020 of CMAT-UM (\url{https://doi.org/10.54499/UIDP/00013/2020}), 
		and by the Portuguese national funds (OE), through the Project FCT/MCTES  PTDC/03091/2022, 
		``Mathematical Modelling of Multi-scale Control Systems: applications to human diseases -- CoSysM3''
		(\url{https://doi.org/10.54499/2022.03091.PTDC}).
		
	\appendix	
	\section{Computations in the weakly nonlinear analysis}
	\label{Appendix}	
	{ We provide here the details of the weakly nonlinear analysis presented in Section \ref{Sec4}, that allows for deeper insight into the formation of patterns in a two-dimensional spatial domain. We start here from the 
	Taylor expansion of system  \eqref{eq:rdsR}-\eqref{eq:rdsE} up to the third
	order around the equilibrium $(M^*,C^*,D^*)$,
	\begin{equation}\label{SistComp1Ap}
		\begin{aligned}
			\frac{\partial {\bf U}}{\partial t} &= \mathcal L\,{\bf U} + \mathcal H[{\bf U}], \quad \mbox{for} \quad
			{\bf U}=\left(\begin{array}{c}U\\V\\W\end{array}\right)=\left(\begin{array}{c}M- M^*\\C-C^*\\D-D^*\end{array}\right),
		\end{aligned}
	\end{equation}
		with
	\begin{equation}
		\mathcal L  ={\mathbb A}+{\mathbb D}\Delta_{\bf x} = \begin{bmatrix}
			j_{11} + d_{11} \Delta_{\bf x}& j_{12}-\xi\,d_{12}\Delta_{\bf x}& j_{13}\\[2mm]
			j_{21} & j_{22} + d_{22}\Delta_{\bf x}& j_{23} \\[2mm]
			j_{31} & j_{32}  & j_{33}
		\end{bmatrix},
	\end{equation}
	 { 
		{where t}he $j_{lm}$ and $d_{lm}$ are the entries of the Jacobian matrix $\mathbb A$  and the diffusion matrix  ${\mathbb D}$    {given by}
		\be \label{sistlin}
		{\mathbb A}=
		\left(
		\begin{array}{ccc}
			\Pi'(M^*)& 0 & 0\\&\\
			\dfrac1\tau & -\dfrac1\tau 	& \dfrac\delta\tau \\&\\
			0& 0&-\kappa\Psi'(M^*)
		\end{array}
		\right),\quad {\mathbb D}= \left(
		\begin{array}{ccc}
			\Phi_0(M^*)& -\xi \Phi_1(M^*)&0 \\[4mm]
			0 & \dfrac \theta\tau  & 0 \\[4mm]
			0 & 0 & 0 
		\end{array}
		\right),
		\ee}
	 { 
		respectively, and $\mathcal H[{\bf U}]=
		\left(
		\mathcal H^1[\bf U],\,
		\mathcal H^2[\bf U],\,
		\mathcal H^3[\bf U] 
		\right)^T ,$
		with
		\begin{equation}					
			\begin{aligned}
				\mathcal H^1[\bf U]=&
				\displaystyle\sum_{i+j+k=2,3} f_{ijk} U^i V^j W^k +\left(l_{01}U+l_{02}U^2\right)\,\Delta_{\bf x}U +\left(m_{01}+m_{02}U\right)\,\nabla_{\bf x} U\cdot\nabla_{\bf x} U \\&
				\quad-\xi\,\left[\left(l_{11}U+l_{12}U^2\right)\,\Delta_{\bf x}V +\left(m_{11}+m_{12}U\right)\,\nabla_{\bf x} U\cdot\nabla_{\bf x} V\right], \\[4mm]
				\mathcal H^2[\bf U]=&\displaystyle\sum_{i+j+k=2,3} g_{ijk} U^i V^j W^k, \\[4mm]
				\mathcal H^3[\bf U]=&\displaystyle\sum_{i+j+k=2,3} h_{ijk} U^i V^j W^k,
			\end{aligned}
		\end{equation}
	}
	where
	\begin{equation}
		\begin{aligned}
			&l_{01}=m_{01}=\Phi_0'(M^*),\quad l_{02}=\frac12\Phi_0''(M^*), &  &\quad m_{02}=\Phi_0''(M^*),&\\[2mm]
			&l_{11}=m_{11}=\Phi_1'(M^*),\quad l_{12}=\frac12\Phi_1''(M^*), &&\quad m_{12}=\Phi_1''(M^*),&\\[2mm]
			&f_{200}=\frac12\Pi''(M^*)=:\tilde f_2, &\quad & f_{300}=\frac16\Pi'''(M^*)=:\tilde f_3,\\[2mm]
			&h_{200}=\frac12 \kappa\Psi''(M^*)(1-D^*)=0, &\quad & h_{101}=-\kappa\Psi'(M^*)=:\tilde h_{11},\\[2mm]
			&h_{300}=\frac16 \kappa \Psi'''(M^*)(1-D^*)=0, &\quad & h_{201}=-\frac12 \kappa \Psi''(M^*)=:\tilde  h_{21},\\
		\end{aligned}
	\end{equation}
	while all the remaining coefficients are zero.

		 Let us now substitute the expansions \eqref{UExp}-\eqref{timeExp} of the vector ${\bf U}$ and of its temporal derivative into the system \eqref{SistComp1Ap} and collect terms at the same order of $\eta$. We obtain three equations as follows:
	\begin{itemize}
		\item[-] order $\eta$:
		\begin{equation}\label{SistOrd1}
			\mathcal L_c
			\left(
			\begin{array}{c}U_1 \\ V_1\\W_1 \end{array}
			\right)
			=0,
			\quad \mbox{ with} \quad
			\mathcal L_c  = \begin{bmatrix}
				j_{11} + d_{11} \Delta_{\bf x}& j_{12}-\xi_c\,d_{12}\Delta_{\bf x}& j_{13}\\[2mm]
				j_{21} & j_{22} + d_{22}\Delta_{\bf x} & j_{23} \\[2mm]
				j_{31} & j_{32}  & j_{33}
			\end{bmatrix},
		\end{equation}
		\item[-] order $\eta^2$:
		\begin{equation}\label{EqOrd2}
			\frac{\pa}{\pa T_1}\left(
			\begin{array}{c}U_1 \\ V_1\\W_1 \end{array}
			\right)=	
			\mathcal L_c
			\left(
			\begin{array}{c}U_2 \\ V_2\\W_2 \end{array}
			\right)
			+ \mathcal H_2\left[\left(
			\begin{array}{c}U_1 \\ V_1\\W_1 \end{array}
			\right)\right],
			\quad \mbox{ with} \quad
			\mathcal H_2\left[\left(
			\begin{array}{c}U_1 \\[2mm] V_1\\[2mm]W_1 \end{array}
			\right)\right]=\left(
			\begin{array}{c}
				\mathcal H_2^1\left[U_1,V_1,W_1\right] \\[2mm]\mathcal H_2^2\left[U_1,V_1,W_1\right]\\[2mm]
				\mathcal H_2^3\left[U_1,V_1,W_1\right] \end{array}
			\right),
		\end{equation}
		\begin{equation}			
			\begin{aligned}
				\mathcal H_2^1\left[U_1,V_1,W_1\right]=&\,l_{01}\, \nabla_{\bf x}\cdot\left(U_1 \nabla_{\bf x} U_1\right)-\xi_c\, l_{11} \nabla_{\bf x}\cdot\left(U_1 \nabla_{\bf x} V_1\right)-\xi_1\,d_{12}\Delta_{\bf x}V_1+\tilde f_2 U_1^2 
				\\[2mm] 	\mathcal H_2^2\left[U_1,V_1,W_1\right]=&\, 0\\[2mm]	\mathcal H_2^3\left[U_1,V_1,W_1\right]=&\,\tilde h_{11} \,U_1\, W_1\end{aligned}
		\end{equation}
		\item[-] order $\eta^3$:
		\begin{equation}\label{EqOrd3}
			\frac{\pa}{\pa T_1}\left(
			\begin{array}{c}U_2 \\ V_2\\W_2 \end{array}
			\right)+	
			\frac{\pa}{\pa T_2}\left(
			\begin{array}{c}U_1 \\ V_1\\W_1 \end{array}
			\right)=	
			\mathcal L_c
			\left(
			\begin{array}{c}U_3 \\ V_3\\W_3 \end{array}
			\right)
			+ \mathcal H_3\left[\left(
			\begin{array}{c}U_1 \\ V_1\\W_1 \end{array}
			\right),\left(
			\begin{array}{c}U_2 \\ V_2\\W_2 \end{array}
			\right)\right],
		\end{equation}
		 { 
			\begin{equation}
				\mathcal H_3\left[\left(
				\begin{array}{c}U_1 \\ V_1\\ W_1 \end{array}
				\right),\left(
				\begin{array}{c}U_2 \\ V_2\\W_2 \end{array}
				\right)\right]=\left(
				\begin{array}{c}
					\mathcal H_3^1\left[U_1,V_1,W_1,U_2,V_2,W_2\right] \\[2mm]\mathcal H_3^2\left[U_1,V_1,W_1,U_2,V_2,W_2\right]\\[2mm]
					\mathcal H_3^3\left[U_1,V_1,W_1,U_2,V_2,W_2\right] \end{array}
				\right),
			\end{equation}
			with
			\begin{equation}			
				\begin{aligned}
					\mathcal H_3^1\left[U_1,V_1,W_1,U_2,V_2,W_2\right]=&\,l_{01}\, \nabla_{\bf x}\cdot\left(U_1 \nabla_{\bf x} U_2+ U_2 \nabla_{\bf x} U_1\right)+l_{02}\, \nabla_{\bf x}\cdot\left(U_1^2 \nabla_{\bf x} U_1 \right)\\[2mm]&-
					\xi_c\left[l_{11}\, \nabla_{\bf x}\cdot\left(U_1 \nabla_{\bf x} V_2+ U_2 \nabla_{\bf x} V_1\right)+l_{12}\, \nabla_{\bf x}\cdot\left(U_1^2 \nabla_{\bf x} V_1 \right)\right]\\[2mm]&-\xi_1\, l_{11} \nabla_{\bf x}\cdot\left(U_1 \nabla_{\bf x} V_1\right)-\xi_1\,d_{12}\Delta_{\bf x}V_2-\xi_2\,d_{12}\Delta_{\bf x}V_1\\[2mm]&+	
					2 \tilde f_2\,U_1\,U_2+\tilde f_3\,U_1^3
					\\[2mm] 	\mathcal H_3^2\left[U_1,V_1,W_1,U_2,V_2,W_2\right]=&\, 0\\[2mm]	\mathcal H_3^3\left[U_1,V_1,W_1,U_2,V_2,W_2\right]=&\, \tilde h_{11} \left(U_1\, W_2 +U_2\,W_1\right)+\tilde h_{21}\,U_1^2\,W_1\end{aligned}
			\end{equation}
		}
	\end{itemize}
	}
	
	{
	Upon solving system \eqref{SistOrd1}, thanks to spectral properties of the operator $\mathcal L_c$, we can write the solution 
	   as in \eqref{Expu1Text}.

	Let us now consider the order $\eta^2$ equation \eqref{EqOrd2}, that can be rewritten as 
	\begin{equation}\label{EqOrd2.2}
		\mathcal L_c
		\left(
		\begin{array}{c}U_2 \\ V_2\\W_2 \end{array}
		\right)
		= 
		\frac{\pa}{\pa T_1}\left(
		\begin{array}{c}U_1 \\ V_1\\W_1 \end{array}
		\right)
		- \mathcal H_2\left[\left(
		\begin{array}{c}U_1 \\ V_1\\W_1 \end{array}
		\right)\right]	.
	\end{equation}
	The subsequent step is to find a solution $(U_2, V_2,W_2)^T$ to system \eqref{EqOrd2.2}.  {  We observe that, in this case, the operator $\mathcal L_c$ can be defined as a linear continuous operator from the Banach space $\mathcal Z=\left(H^2(\Gamma_{\bf x})\right)^3$, which is generated by the functions $e^{i\,{\tilde{\bf k}\cdot\bx}}$, $\tilde{\bf k}\in \mathbb{R}^2$, to $\mathcal X=\left(L^2(\Gamma_{\bf x})\right)^3$, and the inner product is defined as the classical product in $L^2$, see \cite{da2002second,haragus2010local}  for technical details.} The existence of a nontrivial solution of the non-homogeneous problem  \eqref{EqOrd2.2} is guaranteed by the Fredholm solvability condition on the mentioned space   {  that, in general, states the following. Let $B$ be a Fredholm operator on a Banach (or Hilbert) space and $b$ an element in the space; the inhomogeneous equation $Bx = b$ admits at least one solution if and only if $b$ is orthogonal to every element in the kernel of the adjoint operator $B^+$, that is: $Bx = b$ is solvable if and only if  $\left\langle b, v \right\rangle = 0$ for all $v \in \ker(B^+)$.						
	}	
	The condition states that the right-hand side of \eqref{EqOrd2.2} must be orthogonal to the kernel of the adjoint operator of $\mathcal L_c$, say $\mathcal L_c^+$,  {  that is  $$
		\mathcal L_c^+  = \begin{bmatrix}
			j_{11} + d_{11} \Delta_{\bf x}& j_{21}-\xi_c\,d_{12}\Delta_{\bf x}& j_{31}\\[2mm]
			j_{12} & j_{22} + d_{22}\Delta_{\bf x} & j_{32} \\[2mm]
			j_{13} & j_{23}  & j_{33}
		\end{bmatrix},
		$$} and whose kernel is spanned by 
	\begin{equation} { \label{sigmaVec}
			\left(
			\begin{array}{c}1\\\sigma\\0 \end{array}
			\right)e^{ i\,{\bf k\cdot\bx}}+\mbox{c.c.},\quad |{\bf k}|=k_c,\quad}	
		\sigma=\tau \left(-\Pi'(M^*) + \sqrt{-\dfrac{\Pi'(M^*)\, \Phi_0(M^*)}{\theta}} \right),\end{equation}
	{{{   where c.c. denotes the complex conjugate. Once substituting  \eqref{Expu1Text} into the right-hand side of \eqref{EqOrd2.2}, it turns out to be a linear combination of terms  $e^0$, $e^{i\,\bk_j\cdot\bx}$, $e^{2\,i\,\bk_j\cdot\bx}$, $e^{\,i\,(\bk_j-\bk_l)\cdot\bx}$;  let us, then, isolate the coefficients corresponding to  $e^{i\,\bk_j\cdot\bx}$  in the right-hand side of \eqref{EqOrd2.2} defining (with $l,m\neq j$ and $l\neq m$), for $ j=1,2,3,$
				\begin{equation}\label{Rj}
					\begin{aligned}
						\left(
						\begin{array}{c}R_U^j \\[2mm] R_V^j\\[2mm]R_W^j \end{array}
						\right)&=\left(
						\begin{array}{c}\rho\\1\\0 \end{array}
						\right)\frac{\pa \mathcal W_j}{\pa T_1}-\left(
						\begin{array}{c}
							\xi_1\,d_{12}\,k_c^2\,\mathcal W_j+\rho\,r_1\overline{\mathcal W}_l\,\overline{\mathcal W}_m
							\\0\\0  \end{array}
						\right),
					\end{aligned}
				\end{equation}
				with
				\begin{equation}\label{r1}
					r_1=2\,\rho\,\tilde f_2+
					k_c^2\left(-\rho\,l_{01}+\xi_c\,\,l_{11}\right).
				\end{equation}

			The solvability condition implies that $\begin{pmatrix}
				R_U^j,  R_V^j ,  R_W^j
			\end{pmatrix}^T e^{i\,{\bf k_{\it j}\cdot\bx}} $ are orthogonal to \eqref{sigmaVec} and thus\\
			$\left\langle
			\begin{pmatrix}
				R_U^j \\[2mm]  R_V^j \\[2mm]  R_W^j
			\end{pmatrix}
			,
			\begin{pmatrix}
				1 \\[2mm]  \sigma \\[2mm]  0
			\end{pmatrix}\right\rangle 
			= 0$, for $j=1,2,3$, and from \eqref{Rj} we get}
	}
	\begin{equation}
		\label{Solv1}
		\begin{aligned}
			\left(\rho+\sigma\right)\frac{\pa \mathcal W_j}{\pa T_1}=
			\xi_1\,\,k_c^2\,\Phi_1(M^*)\,\mathcal W_j+\left[
			k_c^2\left(-\Phi_0'(M^*)\rho^2+\xi_c\,\Phi_1'(M^*)\rho\right)+ \rho^2\,\Pi''(M^*)\right]\overline{\mathcal W}_l\,\overline{\mathcal W}_m
			, 
		\end{aligned}	
	\end{equation} for $j=1,2,3$.

	Successively,  {  as the right-hand side of \eqref{EqOrd2.2} reads
		\begin{equation}
			\begin{aligned}
				& \left(
				\begin{array}{c}-2\,\rho^2\,\tilde f_2 \\ 0\\0 \end{array}
				\right)\left(|\mathcal W_1|^2+|\mathcal W_2|^2+|\mathcal W_3|^2\right)+\sum_{j=1}^3 \left(
				\begin{array}{c}R_U^j \\[2mm] R_V^j\\[2mm]R_W^j \end{array}
				\right)\,\,e^{i\,\bk_j\cdot\bx}\\
				&\quad+\sum_{j=1}^3 	\left(
				\begin{array}{c}-\,\rho^2\,\tilde f_2+2\,k_c^2\left(\rho^2\,l_{01}-\xi_c\,\rho\,l_{11}\right)
					\\ 0\\0 \end{array}
				\right)\,\mathcal W_j^2\,e^{2\,i\,\bk_j\cdot\bx}\\
				&\quad+\sum_{\substack{j=1,2,3 \\ l\equiv j+1\mbox{ \tiny mod}3}} 	\left(
				\begin{array}{c}-2\,\rho^2\,\tilde f_2+3\,k_c^2\left(\rho^2\,l_{01}-\xi_c\,\rho\,l_{11}\right)
					\\ 0\\0 \end{array}
				\right)\,\mathcal W_j\,\overline{\mathcal W}_l\,e^{\,i\,(\bk_j-\bk_l)\cdot\bx}+\mbox{ c.c.},
			\end{aligned}
		\end{equation}} 
	we look for the solution of \eqref{EqOrd2.2} in the form  given in \eqref{Expu2Text}}
 { and	whose coefficients $X_m, Y_m, Z_m$ can be recovered} by solving the linear equations for the coefficients of $e^0$, $e^{i\,\bk_j\cdot\bx}$, $e^{2\,i\,\bk_j\cdot\bx}$, $e^{\,i\,(\bk_j-\bk_l)\cdot\bx}$ obtained by \eqref{EqOrd2.2}.  {  More specifically, since it holds 
	$$
	\begin{aligned}
		\mathcal L_c\left(
		\begin{array}{c}U_2 \\ V_2\\W_2 \end{array}
		\right)
		=& \mathbb A \left(
		\begin{array}{c}X_0 \\ Y_0\\Z_0 \end{array}
		\right)\left(|\mathcal W_1|^2+|\mathcal W_2|^2+|\mathcal W_3|^2\right)+\sum_{j=1}^3 \left(
		\begin{array}{c}R_U^j \\[2mm] R_V^j\\[2mm]R_W^j \end{array}
		\right)\,\,e^{i\,\bk_j\cdot\bx}\\
		&+\left(A-4k_c^2\mathbb D\right)\sum_{j=1}^3 \left(
		\begin{array}{c}X_{2} \\ Y_{2}\\Z_{2}\end{array}
		\right)\,\mathcal W_j^2\,e^{2\,i\,\bk_j\cdot\bx}\\&+\left(A-3\,k_c^2\mathbb D\right)\sum_{\substack{j=1,2,3 \\ l\equiv j+1\mbox{ \tiny mod}3}} \left(
		\begin{array}{c}X_{1} \\ Y_{1}\\Z_{1}\end{array}
		\right)\,\mathcal W_j\,\overline{\mathcal W}_l\,e^{\,i\,(\bk_j-\bk_l)\cdot\bx}+\mbox{ c.c.},
	\end{aligned}
	$$					
	we have
	\begin{equation}
		\begin{aligned}
			\left(
			\begin{array}{c}X_0 \\ Y_0\\Z_0 \end{array}
			\right)
			=& 
			\left(\mathbb A	\right)^{-1}
			\left(
			\begin{array}{c}-2\,\rho^2\,\tilde f_2 \\ 0\\0 \end{array}
			\right)=
			-\frac{\rho^2\,\Pi''(M^*)}{\Pi'(M^*)}	\left(
			\begin{array}{c}1 \\[2mm] 1 \\[2mm]0 \end{array}
			\right), \\
			\left(
			\begin{array}{c}X_2 \\ Y_2\\Z_2 \end{array}
			\right)=&\left(A-4k_c^2\mathbb D\right)^{-1}
			\left(
			\begin{array}{c}-\,\rho^2\,\tilde f_2+2\,k_c^2\left(\rho^2\,l_{01}-\xi_c\,\rho\,l_{11}\right)
				\\ 0\\0 \end{array}
			\right)\\
			&=\frac{\rho\left(\Pi''(M^*)\,\rho + 4\,k_c^2\left(-\rho\,\Phi_{0}'(M^*) + \xi_c\,\Phi_{1}'(M^*)\right)\right)}{2\left(1 + 4\,k_c^2\,\theta\right)\left(-\Pi'(M^*) + 4\,k_c^2\,\Phi_{0}(M^*)\right) - 8\,k_c^2\,\xi_c\,\Phi_{1}(M^*)}
			\,	\left(
			\begin{array}{c}
				1 + 4\,k_c^2\,\theta\\
				1\\0
			\end{array}
			\right),	\\
			\left(
			\begin{array}{c}X_1 \\ Y_1\\Z_1 \end{array}
			\right)=&\left(A-3k_c^2\mathbb D\right)^{-1}
			\left(
			\begin{array}{c}-2\,\rho^2\,\tilde f_2+3\,k_c^2\left(\rho^2\,l_{01}-\xi_c\,\rho\,l_{11}\right)
				\\ 0\\0 \end{array}
			\right)\\
			&=\frac{\rho\left(\Pi''(M^*)\,\rho + 3\,k_c^2\left(-\rho\,\Phi_{0}'(M^*) + \xi_c\,\Phi_{1}'(M^*)\right)\right)}{\left(1 + 3\,k_c^2\,\theta\right)\left(-\Pi'(M^*) + 3\,k_c^2\,\Phi_{0}(M^*)\right) - 3\,k_c^2\,\xi_c\,\Phi_{1}(M^*)}
			\,	\left(
			\begin{array}{c}
				1 + 3\,k_c^2\,\theta\\
				1\\0
			\end{array}
			\right).
		\end{aligned}
	\end{equation}
	}
}
{
At this point, we pass to the order $\eta^3$ equation \eqref{EqOrd3}, that can be cast in the form
\begin{equation}\label{EqOrd3.2}
\mathcal L_c
\left(
\begin{array}{c}U_3 \\ V_3\\W_3 \end{array}
\right)=
\frac{\pa}{\pa T_1}\left(
\begin{array}{c}U_2 \\ V_2\\W_2 \end{array}
\right)+	
\frac{\pa}{\pa T_2}\left(
\begin{array}{c}U_1 \\ V_1\\W_1 \end{array}
\right)- \mathcal H_3\left[\left(
\begin{array}{c}U_1 \\ V_1\\W_1 \end{array}
\right),\left(
\begin{array}{c}U_2 \\ V_2\\W_2 \end{array}
\right)\right].
\end{equation}
 {  Proceeding as done above, inserting expressions \eqref{Expu1Text} and \eqref{Expu2Text} in \eqref{EqOrd3.2}, along with relations \eqref{Solv1}, we may apply again the Fredholm solvability condition. In this case, the coefficients corresponding to  $e^{i\,\bk_j\cdot\bx}$  in the right-hand side of \eqref{EqOrd3.2} are
\begin{equation}\label{Sj}
	\begin{aligned}
		\left(
		\begin{array}{c}S_U^j \\[2mm] S_V^j\\[2mm]S_W^j \end{array}
		\right)&=\left(
		\begin{array}{c}\rho\\1\\0 \end{array}
		\right)\left(\frac{\pa \mathcal V_j}{\pa T_1}+\frac{\pa \mathcal W_j}{\pa T_2}\right)-\left(
		\begin{array}{c}
			k_c^2\left (\xi_2 d_{12}\,\mathcal W_j + \xi_1\,\left(\mathcal V_j  d_{12} + 
			\rho\, l_{11}\,\overline{\mathcal W}_m\,\overline{\mathcal W}_l\right)\right)
			\\
			+ \rho\,r_1\,\left(\overline{\mathcal V}_l\,\overline{\mathcal W}_m + \overline{\mathcal V}_m\,\overline{\mathcal W}_l\right)\\+ \rho^2\,\left[r_2|\mathcal W_j|^2 + r_3(|\mathcal W_l|^2 + |\mathcal W_m|^2)\right]\,\mathcal W_j
			\\[2mm]0\\[2mm]0  \end{array}
		\right),
	\end{aligned}
\end{equation}
with quantities
\begin{equation}
	\begin{aligned}
		r_1& \mbox{ as} \mbox{ in} \eqref{r1},\\[2mm]
		r_2&=\frac{X_0 + X_2}{\rho} \, 2\tilde f_2  + 3{\tilde f_3 \, \rho}\\[2mm]
		&\quad + k_c^2 \, \left[-\frac{X_0 + X_2}{\rho}  \, l_{01} - {\rho \, l_{02}}+ \xi_c \, \left(\left(\frac{X_0 - X_2}{\rho^2} + 2 \frac{Y_2}{\rho}\right) \, l_{11} + {\, l_{12}}\right)\right],
		\\[2mm]
		r_3&=(X_0 + X_1) \, 2\tilde f_2 \, \rho +  6\tilde f_3 \, \rho^2 \\[2mm]
		&\quad+ 
		k_c^2 \, \left[-\frac{X_0 + X_1}{\rho}\, l_{01} - \rho \, m_{02} +  \xi_c\left(\left(\frac{2 X_0 - X_1}{\rho^2} + 3  \frac{Y_1}{\rho}\right) \, \frac{l_{11}}{2} + { m_{12}}\right)\right],
	\end{aligned}
\end{equation}		
which are recovered by applying $\mathcal H_3$ to the expression of $(U_1,V_1,W_1)^T$ and $(U_2,V_2,W_2)^T$ derived above.
Also in this case, we require 	$\left(
S_U^j, S_V^j, S_W^j\right)^T$ to be orthogonal to the kernel of $\mathcal L_c^+$ given by \eqref{sigmaVec},
} 					
obtaining
\begin{equation}\label{Solv2}
\begin{aligned}
	(\rho + \sigma)\left(\frac{\partial \mathcal V_j}{\partial T_1} + \frac{\partial \mathcal W_j}{\partial T_2}\right) =\, &
	k_c^2\left (\xi_2 \Phi_1(M^*)\,\mathcal W_j + \xi_1\,\left(\mathcal V_j \Phi_1(M^*) + 
	\rho\, \Phi_1'(M^*)\,\overline{\mathcal W}_m\,\overline{\mathcal W}_l\right)\right)
	\\
	&+ r_1\,\left(\overline{\mathcal V}_l\,\overline{\mathcal W}_m + \overline{\mathcal V}_m\,\overline{\mathcal W}_l\right)+ \left[r_2|\mathcal W_j|^2 + r_3(|\mathcal W_l|^2 + |\mathcal W_m|^2)\right]\,\mathcal W_j,
\end{aligned}
\end{equation}
for $j,l,m=1,2,3$, $j\neq l\neq m$.}	
 {Equations \eqref{Solv1} and \eqref{Solv2} provide the multiple scale derivatives included in the evolution equation \eqref{EqAj}. We can outline, then, the equations for  the amplitudes $A_j^U$: 
	\begin{equation}
	\frac{\pa A_j^U}{\pa t}=
	\eta^2\,\,\frac{\pa\mathcal W_j}{{\pa T_1}}+\eta^3\, \,\left(\frac{\pa\mathcal W_j}{{\pa T_2}}+\frac{\pa\mathcal V_j}{{\pa T_1}}\right)+O(\eta^4),
		\end{equation}}
		 {  that omitting higher order terms become
	\begin{equation}\label{SistAj1}
		\begin{aligned}
			\frac{(\rho+\sigma)}{\rho}\frac{\pa A_j^U}{\pa t}=&\textbf{}\eta^2\left[
			\xi_1\,\,k_c^2\,\Phi_1(M^*)\,\mathcal W_j+\rho\,r_1\overline{\mathcal W}_l\,\overline{\mathcal W}_m\right]\\&+\eta^3
			\left[k_c^2	\left(\xi_2 \Phi_1(M^*)\,\mathcal W_j + \xi_1\,	\left(\mathcal V_j \Phi_1(M^*) + 
			\rho\, \Phi_1'(M^*)\,\overline{\mathcal W}_m\,\overline{\mathcal W}_l\right)\right)\right.
			\\
			&\left.+ \rho\,r_1\,\left(\overline{\mathcal V}_l\,\overline{\mathcal W}_m + \overline{\mathcal V}_m\,\overline{\mathcal W}_l\right)+ \rho^2\,\left(r_2|\mathcal W_j|^2 + r_3(|\mathcal W_l|^2 + |\mathcal W_m|^2)\right)\,\mathcal W_j				
			\right],
		\end{aligned}
	\end{equation}
	and can be recast as
	\begin{equation}
		\begin{aligned}
			\frac{(\rho+\sigma)}{\rho}\frac{\pa A_j^U}{\pa t}=&
			k_c^2\,\Phi_1(M^*)\,\left[(\eta^2\xi_1+\eta^3\xi_2)\,\mathcal W_j+\eta^3\xi_1\,\mathcal V_j\right]+\rho\,r_1\,\left[\eta^2\overline{\mathcal W}_l\,\overline{\mathcal W}_m+\eta^3\left(\,\overline{\mathcal V}_l\,\overline{\mathcal W}_m + \overline{\mathcal V}_m\,\overline{\mathcal W}_l\right)\right]\\
			&+\eta^3\, \xi_1\,	\rho\, \Phi_1'(M^*)\,\overline{\mathcal W}_m\,\overline{\mathcal W}_l+ \rho^2\,\left(r_2|\mathcal W_j|^2 + r_3(|\mathcal W_l|^2 + |\mathcal W_m|^2)\right)\,\mathcal W_j.
		\end{aligned}
	\end{equation}	
	From \eqref{expxi} we may write $\eta\, \xi_1=\xi-\xi_c-\eta^2\xi_2 +O(\eta^3)$, obtaining 
	\begin{equation}
		\begin{aligned}
			(\rho+\sigma)\frac{\pa A_j^U}{\pa t}=&
			k_c^2\,\Phi_1(M^*)\,\left[(\xi-\xi_c){\rho}(\eta\,\mathcal W_j+\eta^2\,\mathcal V_j)\right]+\rho^2\,r_1\,\left[\eta^2\overline{\mathcal W}_l\,\overline{\mathcal W}_m+\eta^3\left(\,\overline{\mathcal V}_l\,\overline{\mathcal W}_m + \overline{\mathcal V}_m\,\overline{\mathcal W}_l\right)\right]\\
			&+(\xi-\xi_c)\,	\rho^2\, \Phi_1'(M^*)\,\overline{\mathcal W}_m\,\overline{\mathcal W}_l+\rho^3\, \left(r_2|\mathcal W_j|^2 + r_3(|\mathcal W_l|^2 + |\mathcal W_m|^2)\right)\,\mathcal W_j,
		\end{aligned}
	\end{equation}	
	and, from \eqref{exprAj}, this is equivalent to equation \eqref{SistAj} that we recall here:
\begin{equation}\label{SistAjAp}
	r_0\frac{\pa A_j^U}{\pa t}=\xi_m \,A_j^U+\left(s_1+\xi_m\,\tilde s_1\right)\,\overline {A_l^U}\,\overline{A_m^U}+A_j^U\,\left[s_2\,|{A_j^U}|^2+s_3\left(|{A_l^U}|^2+|A_m^U|^2\right)\right],
\end{equation}
being 
\be r_0=\dfrac{\rho+\sigma}{k_c^2\,\xi_c\,\Phi_1(M^*)},\quad \xi_m=\dfrac{\xi-\xi_c}{\xi_c},\quad \tilde s_1=\dfrac{\rho\,\Phi_1'(M^*)}{\Phi_1(M^*)},\quad s_i=\dfrac{r_i}{k_c^2\,\xi_c\,\Phi_1(M^*)}, \label{CoefAmp}\ee
with $i=1,\,2,\,3$.
}
			
				\bibliographystyle{elsarticle-num} 
				\bibliography{biblio.bib}
				
			\end{document}